\documentclass[twocolumn,eqsecnum,showpacs]{revtex4}

\usepackage{epsfig}
\usepackage{amsmath}

\begin{document}

\title{Renormalized Poincar\'e algebra for effective particles in
quantum field theory}

\author{Stanis{\l}aw D. G{\l}azek}\thanks{
Supported by KBN grant No. 2 P03B 016 18.}
\author{Tomasz Mas{\l}owski}

\affiliation{Institute of Theoretical Physics, Warsaw University,
ul. Ho\.za 69, 00-681 Warsaw, Poland}

\date{October 17, 2001}

\begin{abstract}
Using an expansion in powers of an infinitesimally small coupling
constant  $g$, all generators of the Poincar\'e group in local scalar
quantum field  theory with interaction term $g \phi^3$ are expressed
in terms of annihilation and  creation operators $a_\lambda$ and
$a^\dagger_\lambda$ that result from a  boost-invariant
renormalization  group procedure for effective particles.  The group
parameter $\lambda$  is equal to the momentum-space width of form
factors that appear in  vertices of the effective-particle
Hamiltonians, $H_\lambda$.  It is verified for terms order 1, $g$, and
$g^2$, that the calculated  generators satisfy required commutation
relations for arbitrary values  of $\lambda$. One-particle eigenstates
of $H_\lambda$ are shown to  properly transform under all Poincar\'e
transformations. The transformations  are obtained by exponentiating
the calculated algebra. From a phenomenological  point of view, this
study is a prerequisite to construction of observables  such as spin
and  angular momentum of hadrons in quantum chromodynamics.
\end{abstract}


\maketitle

\section{Introduction}

Complete theoretical description of relativistic particles requires
understanding of their binding, which concerns how these particles
form bound states and how their yet unidentified constituents might be
bound, if any. The fact that potentially relevant binding mechanisms
are still far from being understood is best illustrated by the current
example of  hadrons as built from quarks and gluons. Hadronic
structure  continues to pose problems with  physical interpretation of
phenomena such  as the spin of nucleons or apparent lack of gluonic
excitations in the  spectrum of light hadrons. Phenomenology includes
constituent quarks and  gluons, partons in the infinite momentum
frame, and field-theoretic  concepts of QCD, but no unified
understanding of the building blocks of  hadrons is presently
available.

A verified non-relativistic method for describing bound states in
quantum mechanics is provided by the Schr\"odinger equation 
$H |\psi\rangle = E |\psi\rangle$. However, one  must extend the method
to include at least symmetries of special relativity. In order to
combine quantum mechanics and special relativity in Hamiltonian
dynamics, one needs to construct  ten hermitian generators of the
Poincar\'e algebra \cite{Dirac}, i.e. four momenta  $P^\mu$, $\mu =
0,1,2,3$, and six operators $M^{\mu\nu} = - M^{\nu\mu}$ that generate
rotations and boosts. The ten generators should satisfy following
commutation relations,
\begin{eqnarray}
[P^\mu,P^\nu] & = & 0 \; , \label{pp} \\
{[} P^\mu,M^{\nu\rho}] & = & i(g^{\mu\nu}P^\rho - g^{\mu\rho}P^\nu) \; , 
\label{pm}
\end{eqnarray}
\begin{align}
& {[}M^{\mu\nu}, M^{\rho\sigma}]  \nonumber \\
& = i (g^{\mu\rho} M^{\sigma\nu}-g^{\mu\sigma} M^{\rho\nu} +
g^{\nu\rho} M^{\mu\sigma} - g^{\nu\sigma} M^{\mu\rho}) \; . \label{mm}
\end{align}


In constructing these operators one should certainly include the fact
that relativistic interactions can annihilate and create particles.
This phenomenon can be described using quantum field theory (QFT).
But QFT leads to the problem that when one attempts to calculate the
Poincar\'e generators using local interactions and canonical
quantization, one discovers ultraviolet divergences that have to be
regulated. Unfortunately, regularization of the generators destroys
symmetries of the local theory and Eqs.  (\ref{pm}-\ref{mm}) are not
satisfied. This complication is considered by many authors an
essential drawback of the canonical approach. However,  if some
renormalization group procedure \cite{Wilson12} were able to remove
the regularization dependent terms that violate Poincar\'e symmetry,
the  resulting quantum mechanical approach could become a  candidate
for the relativistic description of particle binding. The desired
renormalization procedure should produce expressions for all
Poincar\'e generators in terms of scale-dependent effective particles,
instead of the bare point-like quanta. A candidate calculus for
development of such effective particle scheme has been recently
proposed \cite{DEG} in the case of Hamiltonian operators, based on the
idea of the similarity renormalization group procedure for Hamiltonian
matrices \cite{GW}. The new calculus is already known to produce
asymptotic freedom in the renormalized Hamiltonians for effective
gluons in QCD. Present work extends the new method to all generators
of the Poincar\'e algebra, using a simplest example of scalar
particles to outline the scheme. Dynamics of particles with spin 1/2
and 1, especially in gauge theories, turn out to involve additional
small-$x$ singularities that require extensive studies before
conclusions can be drawn about applicability of the new method to
gauge theories.

The Poincar\'e generators that do not depend on interactions are
relatively easy to construct in QFT,  since they are equal to the
generators for free particles. They are called  kinematical, since the
transformations they generate correspond to symmetries of the space of
kinematical variables that label states. The generators that involve
interactions  are called dynamical, for they generate transformations
that involve  evolution of states in time, which involves
dynamics. The latter are much  harder to construct in QFT because the
interaction terms one obtains from  local theories involve
divergences. It helps to note at this point that the number of the
hard-to-construct dynamical generators depends  on the form of
dynamics. Three  qualitatively different forms exist  \cite{Dirac}:
the well known equal-time formulation, the point form,  and the front
form, which is conventionally  called here the light-front (LF)
dynamics. These forms differ by a hyper-surface in space-time where
the wave functions for a system under consideration would be
defined. Different surfaces induce  different expressions for the
generators of the Poincar\'e group in QFT.

In the usual formulation, a physical system is defined on the
hyper-plane of equal time, e.g. $t=0$, and the hyper-plane is
invariant under six Poincar\'e transformations. Noether's theorem
implies that there are  six kinematical generators, i.e. three momenta
and three generators of  rotations. The remaining four generators of
translations in time and three  Lorentz boosts, all depend on the
interactions, and they lead to the ultraviolet divergence problems.
In the point form of dynamics, in which the system's evolution is
traced from one space-like hyperboloid $x^2 = const.$ to another,  one
has six kinematical generators, $M^{\mu\nu}$, and four dynamical,
$P^\mu$. This form seems to have a natural relativity structure   but
it encounters difficulties in practice, since all momentum operators
depend on interactions and momentum representation for quantum  states
is not easily available. In the LF dynamics, the system is defined on
the hyper-plane described by equation $x\eta = 0$, where $\eta$ is a
zero-vector, often chosen  to have components $\eta^\mu = (\eta^0=1,
\eta^1=0,  \eta^2=0, \eta^3=-1)$ so that the LF hyper-plane is given
by the condition $x^+=x^0+x^3=0$. There are {\it seven} kinematical
generators in the LF approach:  $P^+$,  $P^\perp$, $M^{+\perp}$, $M^{12}$,
and $M^{+-}$, where $\perp \, = (1,\, 2)$.  The only dynamical
generators are $P^-$, and $M^{-\perp}$.  Moreover,  $P^+$ has
exclusively positive eigenvalues, denoted by $p^+$,  and the free
single particle ``energy'', $p^- = (m^2 + p^{\perp \, 2}) /p^+$, does
not involve a square root.

These features suggest that the construction of dynamical generators
could  be easier in the LF dynamics than in other approaches. Quantum
field theories formulated in the equal-time approach lead to Hamiltonian
terms that can create particles out of the bare vacuum, as
long as the created particles'  three-momenta sum up to zero.  These
terms immediately lead to unknown ground state structures that are
thought to be involved in formation of the effective degrees of
freedom. The corresponding expressions  for effective Poincar\'e
generators can hardly be found before the ground  state and its
excitations are understood. In the regularized LF dynamics, the bare
vacuum state is an eigenstate of all generators with eigenvalue zero,
since creation of particles with positive  momenta $p^+$ would have to
change the state momentum and translation invariance forbids that. The
bare vacuum  state can be then  used in a perturbative approach to
build effective particle states and study their  interactions. These
states and interactions may turn out to contain  very complicated
long-wavelength components that correspond to the vacuum effects in
the standard approach, but the perturbative construction can start
from the physically relevant non-vacuum parts. Thus, the  vacuum effects
in LF dynamics are not expected to overwhelm the bound state problem
from the beginning.

An important argument for considering LF dynamics comes from the
fact that the seventh kinematical generator, i.e. $M^{+-}$, appears to
allow boosting of states to the infinite momentum frame. Therefore, one
may hope that the constituent structure of bound states in the rest frame
of reference can be directly connected with the parton picture in the
infinite momentum frame. A boost-invariant renormalization group procedure 
for effective particles in LF quantum field theory is therefore highly
desired for description of high-energy collisions of particles, especially
hadrons using QCD.

This work is focused on the Poincar\'e algebra in LF dynamics with
simplest interactions that produce ultraviolet divergences, scalar
fields with a cubic interaction term. Key elements of the perturbative
renormalization group calculus for Poincar\'e algebra are related to the
three-prong nature of the interactions. The  same structure is encountered
in all theories of physical interest. Although the scalar theory is
considered unstable, the construction of generators order by order
in perturbation theory never  runs into the need of considering the
exact ground state. Thus, the general renormalization group scheme
can be studied perturbatively using scalar particles. This study exposes
spin-independent features that would be a part of construction of
relativistic LF dynamics of effective particles in all physically
interesting theories, cf. \cite{KS}.

This article is organized as follows. Section \ref{sec2} describes
regularization of the divergent bare canonical generators.  Derivation
of effective generators and verification of their  commutation
relations, are described in Section \ref{sec3}. Section \ref{sec4} shows 
that one particle eigenstates of the Hamiltonian transform under finite
Poincar\'e transformations as they should in a relativistic theory.
An example of dynamical rotation around one of the transverse axes
is presented for illustration. Section \ref{sec5} concludes the paper by a brief
summary of the results and some of their implications.  Four Appendices
contain all details needed for completeness.

\section{Canonical generators}
\label{sec2}

The classical local Lagrangian density

\begin{equation}
{\cal L}=\frac{1}{2}\partial_\mu \phi \partial^\mu \phi -
\frac{1}{2}m^2\phi^2 - \frac{1}{3!} g \phi^3 \; , \label{L}
\end{equation}
\noindent
provides equations of motion for the field $\phi$. Using  Poincar\'e
invariance of $\cal L$, one can introduce \cite{Yan}  the density of
the energy-momentum tensor,

\begin{equation}
{\cal T}^{\mu\nu}= \frac{\partial {\cal L}}{\partial (\partial_\mu
\phi)}  \partial^\nu \phi - g^{\mu\nu} {\cal L} \; ,
\end{equation}
\noindent
and write expressions for $P^\mu$ and $ M^{\mu\nu}$,

\begin{eqnarray}
P^\mu & = & \left. \frac{1}{2} \int d^2x^\perp dx^- {\cal T}^{+\mu}
\right|_{x^+=0} \;  , \label{p} \\
M^{\mu\nu} & = & \left. \frac{1}{2} \int
d^2x^\perp dx^- (x^\mu {\cal T}^{+\nu}  - x^\nu {\cal T}^{+\mu})
\right|_{x^+=0} \; . \label{m}
\end{eqnarray}

At the LF-time $x^+=0$, the field $\phi$ may be decomposed into
Fourier  components,
\begin{equation}
\left. \phi(x)= \int [p] \left( e^{ipx} a_p^\dagger  + e^{-ipx} a_p
\right) \right|_{x^+=0}\; , \label{phi}
\end{equation}
\noindent where $[p] = dp^+ dp^1 dp^2 \theta(p^+)/(16\pi^3 p^+)$.
Imposing commutation relations
\begin{equation}
[a_p,a^\dagger_q]= p^+ \tilde \delta (p -q) \; , \label{comrel}
\end{equation}
\noindent
where $\tilde \delta(k) = 16\pi^3 \delta(k^+)\delta(k^1)\delta(k^2)$,
inserting Eq. (\ref{phi}) into Eqs. (\ref{p}) and (\ref{m}), putting
all creation operators to the left of all annihilation operators, and
dropping all singular terms that result from the ordering, one obtains
a set of heuristic expressions for Poincar\'e generators that are
listed in Appendix \ref{app1}. If the interactions are absent, i.e. $g=0$,
these generators satisfy all commutation relations (\ref{pp}) -
(\ref{mm}).  However, when the interaction is  present, i.e. for $g
\ne 0$, products of $P^-$, $M^{-1}$ and $M^{-2}$ produce divergent
operators. The divergences  result from summation over intermediate
states of unlimited free energies.  To provide meaning to the
otherwise divergent products, one has to impose  some cutoff on the
kinematical momentum variables so that the energy range  in the
summation becomes finite. Introduction of such a cutoff is called
regularization.  In the LF dynamics, the regularization of $P^-$,
$M^{-1}$  and $M^{-2}$ is introduced through artificially inserted
factors $r_\Delta$ that multiply interaction vertices and fall off
to zero when particle momenta change in a single interaction by more
than certain cutoff parameter $\Delta$.

Let $\hat O_n$ denote a coefficient of $g^n$ in the operator $\hat O$,
so that $\hat O = \hat O_0 + g \, \hat O_1  + g^2 \, \hat O_2 + ...$ .
In this convention, the regulated bare generators contain only terms
denoted by $P_{\Delta 1}^-$ and $M_{\Delta 1}^{-j}$. They are written
as follows (see Appendix \ref{app1} for details of the notation).
\begin{eqnarray}
P_{\Delta 1}^- & = & \frac{1}{2} \int[123]  \tilde{\delta} \, r_\Delta
a_1^\dagger a_2^\dagger a_3 + h.c. \; . \\
M_{\Delta1}^{-j} & = & \frac{i}{2} \int[123] \left( \frac{\partial}{\partial
p_3^j}\tilde{\delta} \, \right) r_\Delta a_1^\dagger a_2^\dagger a_3 +
h.c. \; .
\end{eqnarray}
The ultraviolet regularization factors are chosen here in
the form $r_\Delta = \exp(-\kappa^{\perp2}_{12}/ \Delta^2)$, where
$\kappa^\perp _{12}  = (p_1^\perp p_2^+ -p_2^\perp p_1^+)/(p_1^+
+p^+_2)$  is a relative transverse momentum of particles 1 and 2. When
regularization is being removed, $\Delta \rightarrow \infty$ and the
regulating factors $r_\Delta$  tend to 1 for all finite
$\kappa_{12}$. 

The particular choice made here for $r_\Delta$ is not unique and if
the regularization is to be removed, it should not matter how it is
introduced. Nevertheless, in the intermediate steps of deriving the
regularization-independent renormalized theory, one prefers to use
regularizations that make the procedure of removing the cutoff
dependence least complicated. The choice made here for $r_\Delta$ is
dictated  by experience with various other factors and convenience of
using them in calculations. The chosen factor preserves the
kinematical symmetries of LF dynamics and maintains factorization of
longitudinal and transverse momenta. The exponential function is
usefully compatible with analytic integration of the renormalization
group equations. An example of considerations that matter in choosing
$r_\Delta$ is provided by a natural candidate factor in LF dynamics:
$\exp(-{\cal M}^2_{12}/ \Delta^2)$, where ${\cal  M}^2_{12} =
(p_1+p_2)^2$  is a free invariant mass of particles 1 and 2,
squared. That choice introduces $\Delta$-dependence even in tree-like
interaction terms that involve small values of $p_1^+$ or $p_2^+$ for
finite external momenta. This feature is not helpful in the scalar
theory and the present article is simplified by avoiding it,
independently of whether regularization factors that depend on
invariant masses can or cannot help in constructing effective dynamics
in gauge theories where small-$p^+$ singularities in tree-like terms
are common.

Although the regularization renders finite candidates for the
interaction-dependent  generators, the cutoff destroys formal
Poincar\'e symmetry of local  theory. In the model case here, three
Poincar\'e algebra commutation relations are violated by
cutoff-dependent terms of order $g$ and $g^2$. Namely, the commutators
\begin{equation}
B^j_\Delta = [P_\Delta^-,M_\Delta^{-j}] \; , \label{Bj}
\end{equation}
\noindent for $j=1,2$ and
\begin{equation}
B^{12}_\Delta = [M_\Delta^{-1}, M_\Delta^{-2}] \; , \label{B12}
\end{equation}
should be equal to zero in the correct algebra, while the regulating
factors cause that these commutators are not equal to zero. Their
structure is described in Appendix \ref{app1}.

The problem of Poincar\'e symmetry violation is solved in the present
work in perturbation theory up to the terms order $g^2$ using a
renormalization group procedure described in next sections. The
procedure is used to calculate counterterms and derive finite
effective generators in the limit $\Delta \rightarrow \infty$.  The
virtue of the procedure, however, is not merely that the resulting
algebra can be satisfied in perturbation theory, but also that the
resulting effective particle dynamics involves only finite momentum
changes in all interactions. The scale of allowed momentum changes is
determined by the parameter that labels operators derived by solving
the renormalization group equations. Thus, the procedure produces
Poincar\'e generators expressed in terms of effective particles that
are quite different from the bare ones and cannot emit or absorb
momenta comparable with $\Delta$. Note that one could also attempt to
remove the symmetry violating terms of order $g^2$ by adding ad hoc
terms $\tilde{B}^j_\Delta$ to $M_\Delta^{-j}$, with
$[P_{0\Delta}^-,\tilde{B}^j_\Delta]=-B^j_\Delta$, and try to keep
working with bare particles and momentum transfers in interactions
among them ranging up to $\Delta \rightarrow \infty$. This is not what
the renormalization group procedure is about or leads to. For example,
the counterterms one obtains in second order perturbation theory
correct particle masses, instead of adding terms like
$\tilde{B}^j_\Delta$.

\section{Effective generators}
\label{sec3}

The effective Poincar\'e group generators are found using the
renormalization group procedure \cite{DEG} that provides means for
finding counterterms in the initial bare Hamiltonian and defines
annihilation and creation operators for effective  particles,
$a_\lambda$ and $a^\dagger_\lambda$. $\lambda$ is the renormalization
group parameter. In the initial regularized theory $\lambda = \infty$.
Detailed formulas of the effective particle calculus used in this work
are given in Appendices \ref{app2} and \ref{app3}. This Section provides 
a qualitative introduction to the method and extends it to all generators 
of the Poincar\'e algebra. Operators $a_\lambda$ and $a^\dagger_\lambda$ 
are generally denoted by $q_\lambda$ wherever it does not matter which of
the operators is spoken about. The bare particle creation and
annihilation operators, $q_\infty$, are shortly denoted by $q$.

The operators $q_\lambda$ are used to define effective particle basis
states  in the Fock space and to write the Poincar\'e algebra in the
form easy to use in that basis, i.e. in terms of  superpositions of
products of $q_\lambda$.  The generators written in terms of
$q_\lambda$ are also called {\it effective}. They turn out to be free
from the  Poincar\'e symmetry violating cutoff effects in a
perturbative theory in the limit $\Delta \rightarrow \infty$. The
reason, to be explained below, is that they do not directly couple
effective particle states with small and large invariant masses. This
feature is analogous but not identical to removal of large energy
jumps in the original similarity renormalization group procedure
\cite{GW}.  The main difference is that the effective generators are
able to produce larger jumps than the effective time-evolution
generator itself. This feature will be explained later, too.

The required expression for $q = q (q_\lambda)$ is found from
equations that determine renormalized Hamiltonians $P^-_\lambda$. The
initial regularized Hamiltonian $P^-_\Delta$ that includes unknown
counterterms,  is originally expressed in terms of bare operators
$q$. One assumes  that $q$ and $q_\lambda$ are connected by a unitary
transformation that preserves quantum numbers labeling $q$ and
$q_\lambda$,
\begin{equation}
q = {\cal U}_\lambda^\dagger q_\lambda {\cal U}_\lambda \; ,
\label{q}
\end{equation}
and one solves the renormalization group equations from Appendix \ref{app2} to
derive effective Hamiltonians, i.e. $P^-_\lambda$ expressed in terms
of effective operators $q_\lambda$ for all values of $\lambda$.  This
procedure provides ${\cal U}_\lambda$ and Appendix \ref{app3} explains how.

The effective Hamiltonians $P^-_\lambda$ contain vertex form factors
$f_\lambda$ in interaction terms ($\lambda$ is the width of the form
factors  in momentum space). Therefore, the Hamiltonians do not
directly couple  states of effective particles with invariant masses
that differ by more than about $\lambda$ and the interaction terms
need to act at least about $\Delta/\lambda$ times to reach masses
order  $\Delta$ starting from finite masses. Thus, if the effective
Hamiltonian terms for finite momenta have a limit when $\Delta
\rightarrow \infty$, the resulting theory is free from the
regularization dependence to all  orders of perturbation theory.
Therefore, the counterterms required in  $P^-_\Delta$ can be found
from the  condition that for finite relative momenta coefficients of
products of $q_\lambda$ in the operators $P^-_\lambda$ are independent
of $\Delta$ when $\Delta \rightarrow \infty$.  Similar condition is
used for finding counterterms in all generators.

Once the function $q(q_\lambda)$ is found from equations for
$P^-_\lambda$, see Appendix \ref{app3}, all other Poincar\'e generators are
expressed in terms of $q_\lambda$ by a plain substitution. All
divergences in the resulting expressions are identified and
counterterms in the bare generators are introduced as required by the
cancellation of the divergences in the generators expressed in terms
of $q_\lambda$. In the scalar theory, it turns out that only mass
counterterms are required in terms of order $g^2$. Once the
counterterms  in the initial generators are found, one takes the limit
$\Delta \rightarrow  \infty$ in the expressions written in terms of
$q_\lambda$. The resulting operators  constitute the effective
Poincar\'e algebra. The key point of the whole  construction is that
the finite scale $\lambda$ in vertex form factors  destroys the
ability of $n$-th order Poincar\'e generators to change  momenta of
effective particles by more than $n\lambda$ and the limit  $\Delta
\rightarrow \infty$ is well defined in perturbation theory in the
effective Fock space basis.

Since the effective interactions are smoothed by the form factors of
width $\lambda$, the time evolution in physical processes
characterized  by some momentum transfer scale $Q$ is most naturally
described by the effective Hamiltonian $H_Q(q_Q)$.  Namely, using
$H_\lambda(q_\lambda)$ with $\lambda \gg Q$, one would introduce  many
tiny details whose net effect could presumably be absorbed into
renormalization of effective interaction parameters, as in the case of
logarithms of the ratio $\lambda/Q$ that contribute to the running
coupling constant. On the other hand, using $\lambda \ll Q$, one would
have to unnecessarily consider multiple  effective interactions to
build momentum transfers on the order of $Q$  through about
$Q/\lambda$ small steps, each transferring at most about  $\lambda$.
To work with the effective particles with $\lambda = Q$ and to have
control on transformation properties of physical states, it is
desirable  to have all  generators written in terms of $q_Q$.

Mathematically, any bare generator with added counterterms,
$\tilde{A}_\infty(q) =  A_\infty + X_\infty$, is expressed  in terms
of effective  creation and annihilation operators using formula

\begin{equation}
\tilde{A}_\infty [q(q_\lambda)] =  \tilde{A}_\lambda(q_\lambda) \;
. \label{atilde}
\end{equation}

\noindent The symbol ~$\tilde{}$~ is used to indicate that the
regularization cutoff parameter $\Delta$ is still kept finite at this
stage of the calculation.  However, once the counterterms $X_\infty$
are found and the effective generators calculated, one can take the
limit $\Delta \rightarrow \infty$ by putting $r_\Delta=1$ in the
effective generators.  Thus, the effective generators are given by the
formula
\begin{equation}
\left. A_\lambda(q_\lambda)
=\tilde{A}_\lambda(q_\lambda)\right|_{r_\Delta=1} \; . \label{A}
\end{equation}
Results obtained from this formula are described below.

Although operators $q$ expressed in terms of $q_\lambda$ contain terms
of order 1, $g$, $g^2$ and higher, all seven effective kinematical
generators appear unchanged in form, as if they continued to be
independent of interactions.  Namely, the only change that occurs in
them is that the bare creation and  annihilation operators are
transformed into the operators for effective particles of scale
$\lambda$. The coefficients are not changed and they remain equal to
the coefficients found in the interacting canonical theory as well as
in the free theory. This result follows from kinematical symmetries 
of LF dynamics, which are preserved in the renormalization group equations.

On the other hand, the interaction terms in the dynamical generators
are changed considerably and contain products of form factors
$f_\lambda$ that limit momentum transfers by $\lambda$ or $2
\lambda$. The factor of two results from a convolution of two first
order terms, each containing $f_\lambda$.  The appearance of form
factors in this pattern is verified here only in terms of order $g$
and $g^2$. Nevertheless, on the basis of these two cases and observed
regularities that appear along with derivatives over particle momenta
in the generators of rotations, one is compelled to consider it
plausible,  though not proved yet, that in higher orders, say  $n$-th,
the form  factors will appear following the same pattern and limit
momentum transfers  by $n \lambda$. This explains why the generators
of rotations can produce  jumps larger than the jumps allowed by
$P^-_\lambda$ itself.

In the interaction terms order $g$, the only change besides
introduction  of form factors $f_\lambda$ is the replacement of $q$ by
$q_\lambda$.  In terms order $g^2$, new features appear. Namely, there
emerge several additional  interactions with products of more than
three effective operators, in distinction from Eqs. (\ref{p})  and
(\ref{m}) in the bare canonical theory where at most three bare
operators are present.  In addition, terms order $g^2$ involve mass
renormalization. The complete  result, derived using Appendices \ref{app2} 
and \ref{app3}, is following, see also Fig \ref{fig1}. and \ref{fig2}..
\begin{widetext}
\begin{eqnarray}
P^-_{\lambda 1} & = & f_\lambda P^-_1(q \rightarrow q_\lambda) \; ,
\label{pl1} \\
P^-_{\lambda 2} & = & \int [p] \frac{\delta m^2_\lambda}{p^+}
a_{\lambda p}^\dagger a_{\lambda p} \nonumber \\ & & + \int [1234]  [
f_\lambda \, \tilde{\delta} \, V^{22}_\lambda a_{\lambda 1}^\dagger
a_{\lambda 2}^\dagger a_{\lambda 3} a_{\lambda 4}  +  ( f_\lambda \,
\tilde{\delta} \, V^{31}_\lambda \,   a_{\lambda 1}^\dagger a_{\lambda
2}^\dagger a_{\lambda 3}^\dagger a_{\lambda 4} +h.c. ) ] \; ,
\label{pl2}
\end{eqnarray}
\begin{figure}[htb]
\begin{center}
\epsfig{file=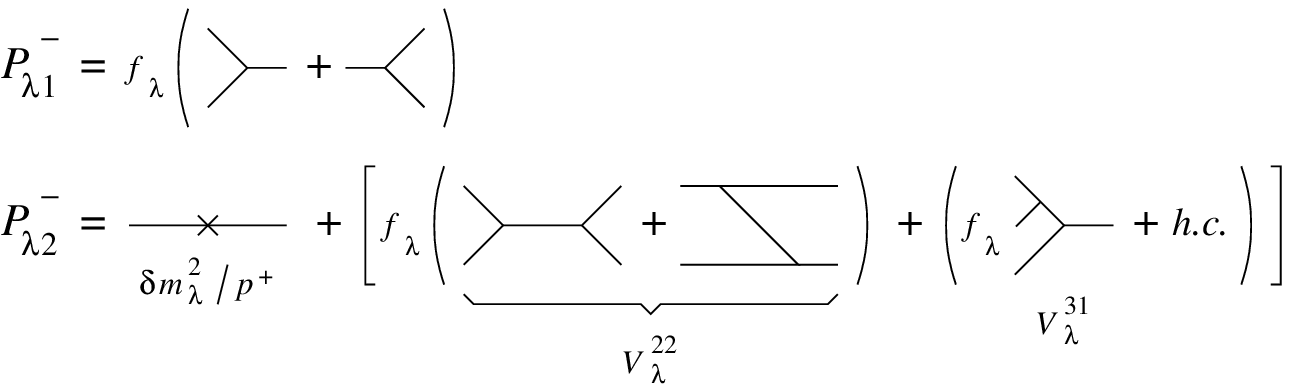}
\end{center}
\caption{Structure of the first and second order interaction terms in
$P^-_\lambda$, Eqs. (\ref{pl1}) and (\ref{pl2}).}
\label{fig1}
\end{figure}
\begin{eqnarray}
M_{\lambda 1}^{-j} & = & f_\lambda M_1^{-j}(q \rightarrow q_\lambda)
\; ,
\label{ml1} \\
M_{\lambda 2}^{-j} & = & i \int [p]\frac{\delta m^2_\lambda}{p^+}
\left( \frac{\partial  a_{\lambda p}^\dagger}{\partial p^j} \right)
a_{\lambda p} \nonumber \\ & & + \; i \int [1234] \frac{1}{p^-_{34} -
p^-_{12}} \left[ f_\lambda D^j (\tilde{\delta} \, V^{22}_\lambda ) +
\tilde{\delta} \, \tilde{V}^{j22}_\lambda \right] a_{\lambda
1}^\dagger a_{\lambda 2}^\dagger  a_{\lambda 3} a_{\lambda 4}
\nonumber \\ & & + \left\{ i \int [1234] \frac{1}{p^-_4 - p^-_{123}}
\left[ f_\lambda D^j (\tilde{\delta} \, V^{31}_\lambda ) +
\tilde{\delta} \, \tilde{V}^{j31}_\lambda \right] a_{\lambda
1}^\dagger a_{\lambda 2}^\dagger a_{\lambda 3}^\dagger a_{\lambda 4}
+ h.c. \right\} \; . \label{ml2}
\end{eqnarray}
\begin{figure}[htb]
\begin{center}
\epsfig{file=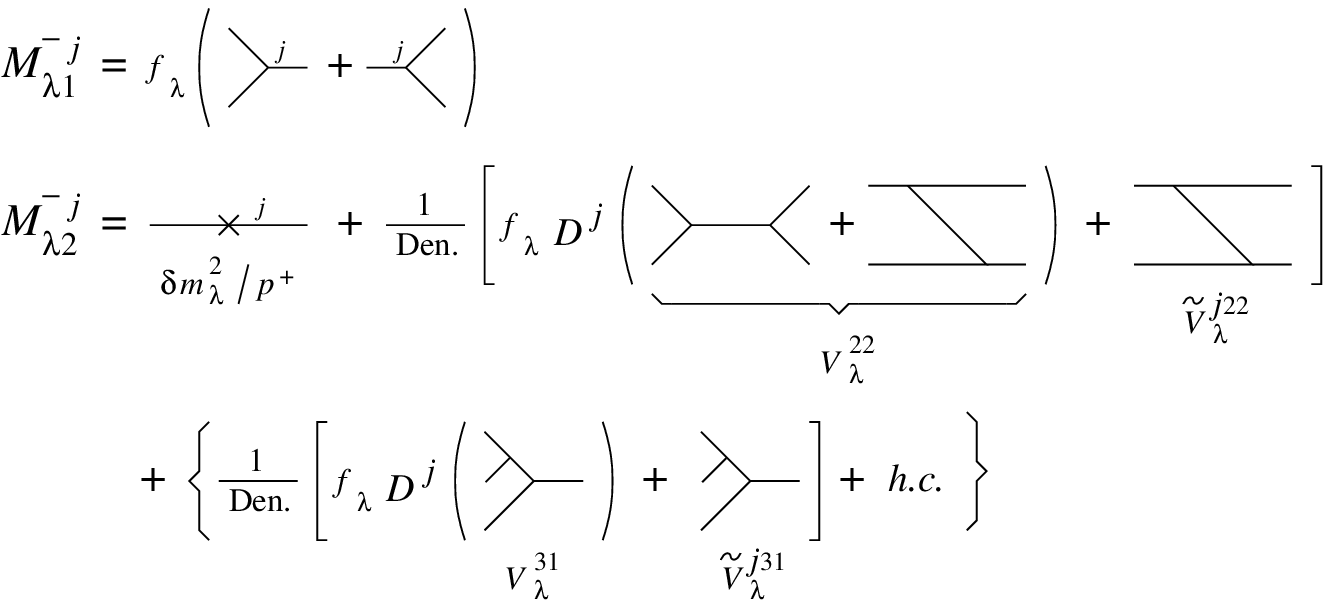}
\end{center}
\caption{Structure of the first and second order interaction terms in
the generators $M_\lambda^{-j}$, $j=1,2$; Eqs. (\ref{ml1}) and
(\ref{ml2}).  The superscript $j$ of $M_\lambda^{-j}$ is carried by
the differential operators on the right-hand side of the figures
(contained also in the functions $\tilde{V}^{j22}_\lambda$ and
$\tilde{V}^{j31}_\lambda$).}
\label{fig2}
\end{figure}
\end{widetext}
The above expressions involve symbols $D^j=\sum_{l=1}^4( p^-_l
\partial / \partial p^j_l + 2 p^j_l \partial / \partial p^+_l)$ and
$p^-_{ij\ldots n} = p^-_i + p^-_j + \cdots + p^-_n$. Symbols
$V^{22}_\lambda$, $\tilde{V}^{j22}_\lambda$, $V^{31}_\lambda$, and
$\tilde{V}^{j31}_\lambda$, are explained in Appendices \ref{app2} 
and \ref{app3}, Eqs. (\ref{V22}), (\ref{tV22}), (\ref{V31}), 
and (\ref{tV31}), respectively.

The terms $\tilde{V}^{j22}_\lambda$ and $\tilde{V}^{j31}_\lambda$ in
$M_{\lambda 2}^{-j}$ have no  counterparts in $P^-_{\lambda 2}$. These
terms turn out to cancel similar terms that appear in the Poincar\'e
algebra commutators, e.g. they cancel  $\lambda$-dependent terms in
$[P^-_{\lambda 1},M_{\lambda 1}^{-j}]$ in the commutator
$[P^-_{\lambda},M_{\lambda}^{-j}]$  evaluated up to order $g^2$. The
standard steps of canonical quantization of local field theory,
regularization, and incorporation of counterterms for $\Delta
\rightarrow \infty$ alone, would not be able to produce the new terms
with finite $\lambda$. The latter are derived here through application
of the renormalization group procedure for effective particles, which
is an additional step in constructing a finite theory.

The renormalized effective mass correction, $\delta m^2_\lambda$  in
$P^-_{\lambda 2}$ and $M_{\lambda 2}^{-j}$, is given by the formula
\begin{equation}
\delta m^2_\lambda = \delta m^2_0 + \frac{1}{32\pi^2} \int_0^1 dx
\int_{{\cal  M}^2_0}^\infty d {\cal  M}^2  \frac{f^2_\lambda
-f^2_{\lambda_0}}{{\cal  M}^2 - m^2}  \; , \label{meff}
\end{equation}
where ${\cal  M}^2_0=m^2/[x(1-x)]$. The free parts of the mass
counterterms in $P^-_{\lambda 2}$ and $M_{\lambda 2}^{-j}$ have to be
chosen equal to one value $\delta m^2_0$, for a chosen $\lambda_0$, to
satisfy the Poincar\'e algebra commutation relations.

The effective generators defined above satisfy all commutation
relations (\ref{pp})-(\ref{mm}) in terms of order 1, $g$, and $g^2$
for arbitrary values of the form factor width $\lambda$. This  result
is checked by explicit calculation. The key features that  allow the
algebra to close for arbitrary finite $\lambda$, even as small as the
mass scale in the theory, are that (1) the form factor $f_\lambda$ is
introduced by integration of renormalization group equations for a
unitary rotation of operators $q$, which preserves commutation
relations,  (2) the argument of $f_\lambda$ is given by squares of
sums of free particle four-momenta that are invariant  under all ten
Poincar\'e transformations with $g=0$, and (3) the initial  condition
from local field theory provides an algebra that is violated  only by
the regularization, which is lifted when one takes the limit $\Delta
\rightarrow \infty$ after calculating the counterterms. It is also 
interesting that the Lorentz symmetry generators come out correctly 
thanks to the preservation of the translational symmetry throughout the
scheme. 

\section{Rotation of physical states}
\label{sec4}

It is well known from earlier works on the effective particle
calculus, that the second-order two-particle scattering matrix  that
results from the LF old-fashioned Hamiltonian approach  using
$P^-_\lambda$, is fully covariant. The only contributions to the
second-order come from tree diagrams with two effective vertices and
from an effective two-body interaction.  These contributions combine
to the same result that is obtained from the Feynman rules. At the
same time no dependence on $\lambda$ appears in the final formula of 
order $g^2$ for scattering amplitudes, precisely as it is required  by 
the renormalization group procedure.  But at such low order of
perturbation theory no mass and no coupling constant renormalization 
are involved in the scattering amplitudes.  The amplitudes that involve
perturbative renormalization group effects are of order $g^4$, and
higher. Therefore, the results obtained in previous  sections for
terms order 1, $g$, and $g^2$, must be extended up to the fourth order
terms to study  symmetries of the scattering matrix calculated in the
effective theory and to verify  applicability of the constructed
interacting algebra to gedanken physical processes with scalar
particles.

However,  even in the second order analysis there exists a test that
the effective algebra has to pass independently of the higher order
terms. Namely, scalar eigenstates of the effective Hamiltonian
$H_\lambda(q_\lambda)$ that are used to construct single-particle
incoming  and outgoing states, should transform accordingly to the
rules of  representation theory for the Poincar\'e group, which state
that \cite{EW}
\begin{equation}
U(\Lambda) |\Psi(p)\rangle = |\Psi(p_\Lambda) \rangle \; . \label{rule}
\end{equation}
This equation means that a unitary operator $U$ representing a
Poincar\'e transformation $\Lambda$ changes a physical scalar
single-particle state of the arbitrary four-momentum $p$ into a
physical single-particle state of the four-momentum $p_\Lambda =
\Lambda p$. This is a testing requirement because $p^-$ component of
the momentum is an eigenvalue of the full interacting Hamiltonian
$H_\lambda(q_\lambda)$, and $U(\Lambda)$ involves interactions through
the interaction-dependent Poincar\'e generators. Note that one has to
know explicit expressions for $U(\Lambda)$ for all Poincar\'e
transformations in order to be able to fully verify symmetry of
observables calculated in effective theories with some value of
$\lambda$. In particular, all physical single-particle states should
actually be labeled only by the values of kinematical variables $p^+$
and $p^\perp$, while the eigenvalue $p^-$ should depend on $p^+$ and
$p^\perp$ as in Eq. (\ref{rowwlas}) below. By construction, this
condition holds for states obtainable from a selected physical
single-particle state using the seven Poincar\'e transformations that
are kinematical in LF dynamics, and by translations in ``time'' $x^+$.
It should also hold for states that result from rotations. Since
rotations are dynamical, they could, in principle, generate anomalies
due to the fact that the LF Hamiltonian approach to quantum field
theory does distinguish a frame of reference in which the quantum
states and operators are defined, and explicit Poincar\'e symmetry is
not kept. In these circumstances, it is interesting to see how the
rotational symmetry of the spectrum is realized in a fixed effective
particle Fock space basis.

In this Section, the physical one-particle states are constructed by
solving  equation $H_\lambda(q_\lambda) |\psi\rangle = E |\psi\rangle$
in  perturbation theory up to terms of order $g^2$. Then, the
generators $A_\lambda(q_\lambda)$ are exponentiated to obtain finite
Poincar\'e group elements and the latter are applied to the
one-particle eigenstates to see if they transform properly, keeping
track of all terms order 1, $g$, and $g^2$. The general procedure is
presented on the example of rotations about one transverse axis. The
construction shows in  detail how the interacting angular momentum and
rotation operators by finite angles act in the LF Fock space. Operators
representing other Poincar\'e transformations are derivable in the
same way and verification of their properties does not need to be
re-produced here explicitly.

Eigenstates of $P^-_\lambda$ with eigenvalues  $p^-_{phys.}=(p^{\perp
2}+m^2_{phys.})/p^+$ are determined by the eigenvalue equation,
\begin{equation}
P^-_{\lambda} |\Psi(p) \rangle = p^-_{phys.}|\Psi(p) \rangle \; .
\label{rowwlas}
\end{equation} Their effective particle Fock space basis
expansion has the form,
\begin{align}
& |\Psi (p) \rangle = N \left[ |\Psi_0 (p) \rangle \right. \nonumber\\
& \left. + g |\Psi_1 (p) \rangle + g ^2 |\Psi_2^{11} (p) \rangle  + g^2
|\Psi_2^2 (p) \rangle + \cdots \right] \; ,\label{state}
\end{align}
where $N$ is a normalization factor, which follows from the condition
\begin{equation}
\langle \Psi (p) | \Psi (p') \rangle = p^+ \tilde \delta (p - p') \; .
\end{equation}
$N$ does not change under Poincar\'e transformations and it factors
out from their analysis. However, $N$ depends on interactions for
given $\lambda$, because it compensates  for the size of components
generated by $P^-_\lambda$. Therefore, if one were interested in
expanding a normalized state in a series of powers of $g$, $N$ would
contribute.

In Eq. (\ref{state}), the state component with subscript 0 denotes
terms with wave-functions order $g^0$, subscript 1 indicates terms
with wave-functions of order $g^1$ with the factor $g$ written
explicitly in front, subscript 2 indicates terms with wave-functions
of order $g^2$ with the factor $g^2$ written explicitly in front, and
so on.   It is necessary to distinguish wave-functions and states in
the perturbative  expansion since the effective particle basis states
depend on interactions and also have an expansion in powers of $g$
when expressed in terms of the bare particle basis states. Since the
latter expansion is already included in the effective creation
operators $a^\dagger_\lambda$, the expansion for the wave-functions in
the effective basis is separated.  This separation is also related to
the fact that in higher order analysis, including renormalization
group evolution of the coupling constant $g_\lambda$ that starts with
terms of order $g^3$, the wave-functions would be expanded in a series
of powers of $g_\lambda$, or found numerically in a non-perturbative
calculation using  $H_\lambda(q_\lambda)$. In the present work that
includes only terms up  to order $g^2$, no difference between $g$ and
$g_\lambda$ is visible.   However, the expansion  of wave-functions
should be thought about as   expansion in a series of powers of
$g_\lambda$, not the initial $g$.

The superscripts in Eq. (\ref{state}) indicate the origin of terms in
perturbation theory using $H_\lambda(q_\lambda)$.  Terms with
superscript 11 result from double action of Hamiltonian terms order
$g$ on the component $|\Psi_0\rangle$, i.e. they originate from the
second order perturbation theory for wave-functions in the effective
dynamics of scale $\lambda$. In contrast, terms with superscript 2
result from single action of second order terms in the effective
Hamiltonian  of scale $\lambda$. In other words, the latter terms
should be thought about as proportional to $g_\lambda^2$ and coming
from a single action of a second-order interaction term from
$H_\lambda(q_\lambda)$ on the component $|\Psi_0\rangle$, i.e. they
originate from the first order perturbation theory for wave-functions
in the effective dynamics of scale $\lambda$.  As further example of
the superscript notation, the component $|\Psi_1\rangle$ could carry a
superscript 1, which is omitted.

Direct calculation gives the following results for physical
one-particle states, starting from one-effective-particle state,
\begin{equation}
|\Psi_0(p) \rangle = a^\dagger_{\lambda p} |0 \rangle \; .
\end{equation}
First order perturbation theory gives terms $|\Psi_1 \rangle$ that
contribute in order $g$,
\begin{equation}
|\Psi_1(p) \rangle = \frac{1}{2} \int [12] \tilde{\delta} (p-p_{12})
\frac{f_\lambda}{p^- - p^-_{12}} |12\rangle,
\end{equation}
where $|12\rangle = a^\dagger_{\lambda 1} a^\dagger_{\lambda 2}
|0\rangle$, and terms $|\Psi_2^2 \rangle$ that contribute in order
$g^2$,
\begin{equation}
|\Psi_2^{2}(p) \rangle = \int [123]\tilde{\delta}(p - p_{123})
\frac{f_\lambda}{p^- -p^-_{123}} V^{31}_\lambda |123 \rangle \; ,
\end{equation}
where $|123\rangle = a^\dagger_{\lambda 1} a^\dagger_{\lambda 2}
a^\dagger_{\lambda 3} |0\rangle$. Second order perturbation  theory
gives $|\Psi_2^{11} \rangle$ that contributes in order $g^2$,
\begin{equation}
|\Psi_2^{11}(p) \rangle = \int [123]  \frac{\tilde{\delta}(p -
p_{123})f_{12}f_{(12)3}} {2p^+_{12}(p^- - p^-_{123})(p^- -
p^-_{(12)3})} |123\rangle \; .
\end{equation}
Here, $p^-_{(12)3}$ stands for $(p^{\perp \,
2}_{12}+m^2)/p^+_{12}+p^-_3$, which means that the total
three-momentum of particles 1 and 2, $p^\perp_{12}$ and  $p^+_{12}$,
is turned into a one-particle momentum and the corresponding $p^-$ is
evaluated using mass $m$. The same convention for $p^-_{(12)3}$  is
used in evaluating $f_{(12)3}$. The subscripts $ij$ in $f_{ij}$, with
$(12)$ understood  as one particle subscript such as $i$, indicate
that one of the invariant masses in the definition (\ref{deff}) is
given by $(p_i + p_j)^2$, while the other by $m^2$. The eigenvalue
Eq. (\ref{rowwlas}) leads to, cf. Eq. (\ref{dtm}),
\begin{widetext}
\begin{equation}
m^2_{phys.} =  m^2 + g^2 \delta m^2_0 + \frac{g^2}{32\pi^2} \int_0^1
dx  \int_{{\cal M}^2_0}^\infty d {\cal  M}^2
\frac{f^2_{\lambda_0}}{{\cal  M}^2 - m^2}  = m^2 + g^2 \delta
\tilde{m}^2_0 \; . \label{mphys}
\end{equation}
\end{widetext}
This result shows that $m^2_{phys.}$ is independent of $\lambda$, as
required  for a physical quantity.

The next step is to show that effective Poincar\'e transformations
obtained by exponentiation of the algebra derived in Section \ref{sec3},
properly transform single physical particle states up to terms of
order $g^2$ in the perturbative series. This is done below on the
example of rotations around transverse axis number 1. As already
mentioned, other rotations and all other Poincar\'e transformations
are verified term by term in the series expansion in powers of $g$
using the same scheme.

The effective generator of rotations around the transverse axis number
1 is given by
\begin{equation}
J^1_\lambda = \frac{1}{2} (M^{-2}_\lambda - M^{+2}_\lambda ) \; ,
\end{equation}
where $M^{-2}_\lambda$ depends on interactions, see Eqs. (\ref{ml1})
and (\ref{ml2}). Physically, finite rotations around axis number 1 by
an arbitrary angle $\alpha$ change the standard spatial three-momenta,
$\vec p_{phys.} = (p^1, p^2, p^3)$, so that
\begin{widetext}
\begin{equation}
(p^1,p^2,p^3) \longrightarrow (p^1, p^2 \cos \alpha - p^3 \sin \alpha,
p^3 \cos \alpha +  p^2 \sin \alpha )\, =: \, \vec p_{phys.\,\alpha }
\; ,
\label{obrot}
\end{equation}
\end{widetext}
and $p^0_{phys.}$ remains unchanged, so that the condition that
$p^2_{phys.} = m^2_{phys.}$ is preserved: $p^2_{phys.\,\alpha } =
m^2_{phys.}$. Note that $p^3=(p^+ - p^-_{phys.})/2$.  The quantum
operator that represents this finite rotation is given by
\begin{equation}
U(\alpha)=e^{-i \alpha J^1_\lambda} \; .
\end{equation}
One verifies the rule (\ref{rule}) by checking if the state
\begin{equation}
|\Psi_\alpha(p) \rangle := U(\alpha) |\Psi(p) \rangle \; ,
\end{equation}
whose expansion in perturbation theory follows from the  expression
\begin{eqnarray}
& |\Psi_\alpha(p) \rangle = N[U_0(\alpha) + g U_1(\alpha) + g ^2
U_2(\alpha) + o(g^3)] \nonumber \\ 
& \times [|\Psi_0(p) \rangle + g|\Psi_1(p) \rangle 
+ g^2 |\Psi_2(p)\rangle + o(g^3) ] \; , 
\end{eqnarray}
satisfies the relation
\begin{equation}
|\Psi_\alpha(p) \rangle = |\Psi(p_{phys. \, \alpha}) \rangle \;
 .\label{tpop}
\end{equation}
Details of the calculation are given in Appendix \ref{app4}. The results can be
written in a concise form by introducing a new symbol $p_{\alpha}$ to
denote the rotated momentum that has the same components $p^+$ and
$p^\perp$ as $p_{phys.}$ before the rotation, while the rotated
components $p^+_\alpha$ and $p^\perp_\alpha $ are calculated by
applying the rotation by angle $\alpha$ to the four-momentum with
the minus component given by condition $p^- = (p^{\perp 2}+m^2)/p^+$,
i.e. the condition $p^2 = m^2$ instead of $p^2 = m^2_{phys.}$. Thus,
$p_{\alpha}$ is obtained from the same relation (\ref{obrot}) that
defines $p_{phys. \, \alpha}$, but with $m^2_{phys.}$ replaced by
$m^2$, see Eq. (\ref{p0}). One obtains,
\begin{eqnarray}
|{\Psi_\alpha}_0 (p) \rangle & = & |\Psi_0 (p_{\alpha}) \rangle \; ,
\label{pa0} \\
|{\Psi_\alpha}_1 (p) \rangle & = & |\Psi_1 (p_{\alpha}) \rangle \; ,
\label{pa1}
\end{eqnarray}
\begin{align}
& |{\Psi_\alpha}_2 (p) \rangle = |\Psi_2 (p_{\alpha}) \rangle \nonumber\\ 
& + \; \frac{\delta m^2_\lambda - \delta \tilde{m}^2_\lambda}
{p^+_{\alpha}} \left( \frac{\sin \alpha}{2} \frac{\partial}{\partial
p^2} - \sin^2 \frac{\alpha}{2} \frac{\partial}{\partial p^+} \right)
a^\dagger_{\lambda p_{\alpha}} | 0 \rangle \; , \label{pa2}
\end{align}
where $\delta \tilde{m}^2_\lambda$ is given by Eq. (\ref{dtm}).

\begin{figure}[htb]
\begin{center}
\epsfig{file=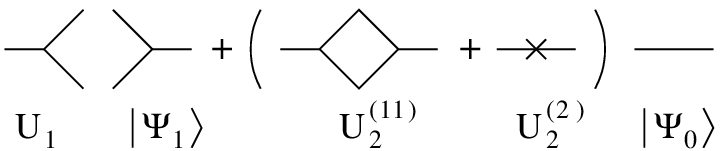}
\end{center}
\caption{Graphical representation of terms that contribute  to the
dynamical mass effects in rotations, see Eq. (\ref{pa2}) and Appendix
\ref{sot}.}
\label{fig3}
\end{figure}

The change of $p$ to $p_{\alpha}$ in the rotated components in Eqs.
(\ref{pa0}) and (\ref{pa1}), agrees with the rule (\ref{rule}). 
Eq. (\ref{pa2}) also agrees with that rule but the second term in 
$|{\Psi_\alpha}_2 (p) \rangle$ on the right-hand side of Eq. (\ref{pa2}) 
requires explanation. It comes from terms that are shown graphically 
in Fig. \ref{fig3}.
\noindent
To check that Eq. (\ref{pa2}) is correct, one needs to replace $p$  in
$|\Psi_0 (p) \rangle$ with $p_{phys. \, \alpha}$.  When the correction
order $g^2$ in the physical mass, i.e. $g^2 \delta \tilde{m}^2_0$, is
small, one can calculate $a^\dagger_{p_{phys. \, \alpha}}$ in the
vicinity of $a^\dagger_{p_{\alpha}}$ and neglect all terms of order
higher then $g^2$. The result is
\begin{align}
& a^\dagger_{\lambda p_{phys. \alpha}} 
= a^\dagger_{\lambda p_{\alpha}} \nonumber\\ 
& + g^2 \; \frac{\delta \tilde{m}^2_0 }{p^+_{\alpha}}
\left( \frac{\sin \alpha}{2} \frac{\partial}{\partial p^2} - \sin^2
\frac{\alpha}{2} \frac{\partial}{\partial p^+} \right)
a^\dagger_{\lambda p_{\alpha}} \; . \label{pa2a}
\end{align}
The operator correction of order $g^2$ in Eq. (\ref{pa2a}) compensates
the operator that acts on the vacuum state, $|0\rangle$, in Eq.
(\ref{pa2}).  Since the state components $|\Psi_1\rangle$ and $|\Psi_2
\rangle$ have wave-functions of order $g$ and $g^2$, one can simply
replace $p_{phys.}$ with $p$  and $p_{phys. \, \alpha}$ with
$p_{\alpha}$ in these terms  without any change in the wave-functions.
However, in the term with wave-function order $g^0=1$, the difference
between $p_{\alpha}$ and $p_{phys. \, \alpha}$ in rotated effective
creation operators does matter for terms of order $g^2$ in the rotated
eigenstate. The second order difference turns out to be entirely
contained in the mass correction. This difference  is correctly
restored by the exponentiated generator $J^1_\lambda$ through
contractions with two particle components of the physical states.  The
origin of the second term on the right-hand side of Eq. (\ref{pa2}) is
thus explained.

In summary, the derived angular momentum operator $J^1_\lambda$
generates correct rotations of single physical particle states in all
terms of second order perturbation theory. In the same way one
calculates expressions for other quantum Poincar\'e transformations
and verifies with the same accuracy that they transform one particle
states in agreement with requirements of special relativity.

\section{Conclusion}
\label{sec5}

The Poincar\'e group generators obtained from scalar quantum field
theory in second-order renormalization group procedure for effective
particles, fulfill all Poincar\'e algebra commutation relations in
terms of order 1, $g$ and $g^2$. For example, the algebra
includes the hard-to-satisfy relation in LF dynamics,
\begin{equation}
[J^i_\lambda, J^j_\lambda] = i \epsilon_{ijk} J^k_\lambda + o(g^3) \; .
\end{equation}
This relation was not easy to derive in LF dynamics because two angular 
momentum operators depend on interactions. On the other hand, the 
dynamical parts of $J^i_\lambda$, $i=1, 2$, are necessary to produce 
the correct mass values and corresponding terms in rotated states. 
The full Poincar\'e algebra is satisfied with the same accuracy for 
arbitrary values of the renormalization group parameter $\lambda$.

Since $\lambda$ equals to the width of vertex form factors in the
renormalized Hamiltonians $H_\lambda (q_\lambda)$, the construction
described in this work links a relativistic local field  theory with a
smooth quantum mechanics of effective particles in the  Fock
space. For $\lambda=\mu$ that matches a physically relevant scale,
observable quantities are expected to be most conveniently calculable
from the Schr\"odinger equations with corresponding $H_\mu (q_\mu)$,
using computers that need to handle only  a limited set of basis
states required to cover the range of momentum scales around
$\mu$. The new result of this work is that the Poincar\'e symmetry  of
these observables can be sought in a sequence of well defined
approximations based on the renormalization group procedure for
effective particles, too. The key remaining question of the scalar
example, studied here only up to terms of order $g^2$, is whether it
can indeed be systematically improved as expected by including terms
of order $g^n_\lambda$ with $n > 2$. The difference between series
expansion in powers of $g$ and expansion in powers of $g_\lambda$,
will appear first in terms of order $g^3$.

Exponentiation of the effective algebra gives operators that represent
finite Poincar\'e transformations. They all turn out to change
physical single-particle states in agreement with requirements of
special relativity. This result includes spatial rotations in the LF
dynamics and was also verified up to terms of order $g^2$. No
anomalies  were found. All non-canonical terms obtained from solutions
of renormalization group equations had to be included in the
Hamiltonian of finite $\lambda$ and in the corresponding angular
momentum operators to  obtain this result. It is natural to expect
that similar but more complicated  terms have to be included in
effective theories of quarks and gluons when one tries to interpret
data for scale-dependent hadronic observables, such as spin structure
of the proton \cite{Hari}, in terms of constituents. It is therefore 
certainly interesting to carry out bound state studies using second 
order effective Hamiltonians in the scalar theory, since they can 
provide basic intuition about interaction-dependent effects in 
the angular momentum of partons. However, the basic study must be 
considerably extended to incorporate spin of individual particles 
in order to become directly applicable to LF QCD. The extension  
requires solution to problems with additional small-$p^+$ singularities  
that characterize gauge bosons \cite{Wilsonetal}. The latter problem  
has to be solved in order to explicitly construct boosted and rotated  
hadronic bound states in the Fock space of effective quarks and gluons.  
Such construction could help, for example, in interpreting relativistic  
partial wave analysis of decays of exotic hybrid states.

Although the original perturbative analysis of hadronic processes with
high momentum  transfers \cite{LB} raises hopes for a constituent
picture of hadrons to be valid in the available range of energies, one
has to deal with the fact that at the low energy end initial bound
state studies \cite{BPA} in LF QCD indicate that spin multiplets  of
excited states in spectra of approximate effective Hamiltonians may
contain splittings that violate rotational symmetry and demand
understanding. If effective generators of rotations in LF QCD with
small-$\lambda$ were available, one  could transform members of a
broken multiplet and determine what kind of components are
generated. The latter could be compared with the structure of
eigenstates in  the broken multiplet. Such comparison could help in
finding out what effective Fock components are missing in the
approximate calculations.

\appendix

\section{Bare Poincar\'e generators}
\label{app1}

LF co-ordinates are defined to be $x^\pm = x^0 \pm x^3$ and $x^\perp=
(x^1, \; x^2)$, with a scalar product of two four-vectors equal
$ab=a^+b^-/2 + a^-b^+/2 - a^\perp b^\perp$. Definitions (\ref{p}) and
(\ref{m}) lead to the following kinematical generators,

\begin{eqnarray}
P^+ & = & \int [p]p^+ a_p^\dagger a_p\; , \\ P^i & = & \int [p] p^i
a_p^\dagger a_p\; , \\ M^{+-} & = & -2i\int [p]p^+ \frac{\partial
a_p^\dagger }{\partial p^+}a_p \; , \\ M^{+j} & = & i\int [p]p^+
\frac{\partial a_p^\dagger } {\partial p^j}a_p\; , \\ M^{12} & = &
i\int [p]\left( p^1 \frac{\partial a_p^\dagger} {\partial p^2}a_p -p^2
\frac{\partial a_p^\dagger }{\partial p^1}a_p \right) \; ,
\end{eqnarray}
and dynamical generators,
\begin{widetext}
\begin{eqnarray}
P^- & = & P^-_0+ g P^-_1= \int [p] p^- a_p^\dagger a_p
+\frac{1}{2}g\int [123]\tilde{\delta} ( a_1^\dagger a_2^\dagger a_3  +
h.c.)\; , \label{p-} \\ M^{-j} & = & M^{-j}_0+ g M^{-j}_1 \nonumber \\
& =  & i\int [p]\left( p^- \frac{\partial a_p^\dagger }{\partial p^j}
a_p + 2p^j \frac{\partial a_p^\dagger }{\partial p^+}a_p \right) +
\left[ \frac{i}{2}g\int [123] \left( \frac{\partial }{\partial p_3^j}
\tilde{\delta} \right) a_1^\dagger a_2^\dagger a_3 + h.c. \right] \;
. \label{m-}
\end{eqnarray}
\end{widetext}

\noindent The symbol $\tilde \delta$, which appears also in
Eq. (\ref{comrel}),  always denotes the three-momentum conservation
Dirac $\delta$-function times  $16\pi^3$. For example, in
Eq. (\ref{m-}),  $\tilde \delta = 16\pi^3 \delta (p_3^+ -
p_{12}^+)\delta(p_3^1 - p_{12}^1) \delta(p_3^2 - p_{12}^2)$, where
$p_{12}^{+,\perp} = p_1^{+,\perp} + p_2^{+,\perp}$.   The arguments of
$\tilde{\delta}$  are always given by  $\sum_{i \in out} p_i - \sum_{j
\in in} p_j$, where the momenta in  the set $in$ are always those of
particles annihilated in the  interaction and momenta in the set $out$
are always those of particles  created in  the interaction. Arguments
of $\tilde{\delta}$ are  written explicitly only if needed. The symbol
$[p]$ denotes integration measure,
\begin{equation}
[p]=\frac{d^2p^\perp dp^+}{16\pi^3 p^+} \theta (p^+) \; .
\end{equation}

The first order term in the regularization-dependent commutator in
Eq. (\ref{Bj}), is
\begin{widetext}
\begin{equation}
B^j_{\Delta 1}= \frac{i}{2} \int [123]   \left[\sum_{l=1}^2 \left(
p^-_l \frac{\partial }{\partial p^j_l} + 2p^j_l \frac{\partial
}{\partial p^+_l} \right) r_\Delta (12) \right] \tilde{\delta} \;
a_1^\dagger a_2^\dagger a_3 - h.c.\; ,
\end{equation}
and the second order term is
\begin{eqnarray}
B^{j}_{\Delta 2 } & = & \left. i \int[1234]
\frac{\theta(p^+_1-p^+_3)}{p^+_1-p^+_3} \left\{
\frac{\partial}{\partial p^j} \Big[ r_\Delta (2p) r_\Delta (3p)
\Big] \right|_{p=p_1-p_3}\right\} \tilde{\delta} \; 
a_1^\dagger a_2^\dagger a_3 a_4
\nonumber \\
& & + \left\{ \frac{i}{2} \int [1234] \frac{1}{p^+_{12}} r_\Delta (12)
\left[ \left. \frac{\partial}{\partial p^j} r_\Delta (3p)\right|_{p=p_1+p_2}
\right] \tilde{\delta} \; a_1^\dagger a_2^\dagger a_3^\dagger a_4  -
h.c. \right\} \; .
\end{eqnarray}
\end{widetext}
The commutator $B^{12}_\Delta$ of Eq. (\ref{B12}), has a similar
structure to $B^j_\Delta$. All these terms would vanish if the
regularization factors were absent.

\section{Renormalization group for particles}
\label{app2}

Starting from the regularized Hamiltonian $H_\Delta$ with counterterms,
one constructs a family of annihilation and creation operators, parametrized
by $\lambda$. Operators $a$ and $a^\dagger$ in $H_\Delta$ correspond
to $\lambda=\infty$. When $H_\Delta$ is
re-written in terms  of $a_\lambda$ and $a^\dagger_\lambda$, the
coefficients in front of products of these operators depend on
$\lambda$, but the Hamiltonian remains unchanged, which is expressed
as $H_\lambda(q_\lambda) =  H_\Delta(q)$. Thus, all Hamiltonians with
different $\lambda$s are equal  and for any $\lambda_1$ and $\lambda_2$,
\begin{equation}
H_{\lambda_1} (q_{\lambda_1}) = H_{\lambda_2} (q_{\lambda_2}) \; .
\label{eqh}
\end{equation}
Introducing ${\cal H}_\lambda = H_\lambda (q)$, which has the same
coefficients in front of $q$s as $H_\lambda(q_\lambda)$ has in front
of $q_\lambda$s, using the equality of Hamiltonians (\ref{eqh}) and
unitary relation (\ref{q}), one obtains
\begin{equation}
{\cal H}_\lambda = {\cal U}_\lambda^{\dagger} {\cal H}_\infty {\cal
U}_\lambda,
\label{UHU}
\end{equation}
\noindent where all operators are written in terms of constant,
i.e. independent of $\lambda$, bare operators $q$. $H_\lambda (
q_\lambda)$  is obtained from  ${\cal H}_\lambda$ by replacing
$q \equiv q_\infty$ with $q_\lambda$.

Since $H_\lambda$ is to contain vertex form facthors that limit changes
of kinetic energy of interacting effective particles, one assumes
$H_\lambda$ to have the form
\begin{equation}
H_\lambda = F_\lambda[G_\lambda] \; .
\end{equation}
\noindent $F_\lambda$ is a linear operation that changes any operator
$\hat{O}_\lambda$,
\begin{widetext}
\begin{equation}
\hat{O}_\lambda =\int[1..n\,n+1\,..m] \tilde{\delta} \, v(1,..,m) \,
 a_{\lambda 1}^\dagger \cdots a^\dagger_{\lambda n}
a_{\lambda \, n+1} \cdots a_{\lambda m} \; ,
\end{equation}
\noindent by introducing the form factor $f_\lambda$,
\begin{equation}
F_\lambda \hat{O}_\lambda =\int[1..n\,n+1\,..m]
\tilde{\delta} \, f_\lambda \, v(1,..,m) \,
a_{\lambda 1}^\dagger \cdots a^\dagger_{\lambda n}
a_{\lambda \, n+1} \cdots a_{\lambda m} \; ,
\end{equation}
\end{widetext}
\noindent where
\begin{equation}
f_\lambda = \exp \left[ -  \frac{({\cal  M}^2_{in}-{\cal
M}^2_{out})^2}{\lambda^4} \right] \; , \label{deff}
\end{equation}
\noindent ${\cal M}^2_{in} = (p_1 + ... + p_n)^2$, and ${\cal
M}^2_{out} = (p_{n+1} + ... + p_m)^2$.  In an abbreviated
notation, one writes $H = fG$. One introduces then  ${\cal G}_\lambda$,
which is connected with $G_\lambda$ in the same way as ${\cal
H}_\lambda$ is with $H_\lambda$.  ${\cal G}_\lambda$ is divided into
two parts,  ${\cal G}_\lambda = {\cal G}_0 + {\cal G}_{I\lambda}$,
where  ${\cal G}_0={\cal G}_\lambda(g=0)$. ${\cal G}_{I\lambda}$ is
assumed to satisfy the  following differential equation (see
e.g. \cite{DEG}),
\begin{equation}
\frac{d}{d\lambda}{\cal G}_{I\lambda}  = \left[ f{\cal G}_{I\lambda},
\left\{ \frac{d}{d\lambda}(1-f){\cal G}_{I\lambda} \right\}_{{\cal G}_0}
\right] \; ,\label{dG}
\end{equation}
where for any operator $\hat O$, $\hat Q = \{ \hat O \}_{{\cal G}_0}$
indicates a solution of equation $[\hat Q, {\cal G}_0]= \hat O$.
Equation (\ref{dG})
guarantees that perturbation theory is free from small energy
denominators, effective interactions are connected, they possess
required  cluster properties and preserve all kinematical symmetries
of LF dynamics. To solve Eq. (\ref{dG}) with accuracy to terms of
order $g^2$,   one writes
\begin{equation}
{\cal G}_{I\lambda} = g \tau_1 + g^2 \tau_2 +\cdots \; .
\end{equation}
Equation (\ref{dG}) implies
\begin{eqnarray}
\tau_1' & = & 0  \; , \label{tau1} \\
\tau_2' & = & [\{f' \tau_1\},f \tau_1]
= (\{f'\}f-f\{f'\}) [\tau_1\tau_1 ]\; . \label{tau2}
\end{eqnarray}
Integrating (\ref{tau1}), one obtains $\tau_{\lambda 1}=\tau_{\infty
1}$, which means that
\begin{equation}
P^{-}_{1\lambda }=f_\lambda \tau_{\lambda 1}(q\rightarrow q_\lambda)
= \frac{1}{2} \int [123]\tilde{\delta} \,
f_\lambda a_{\lambda 1}^\dagger a_{\lambda 2}^\dagger a_{\lambda 3} + h.c. \; .
\end{equation}
In contrast to (\ref{p-}), $a^\dagger$ and $a$ in the effective
Hamiltonian  correspond to the scale $\lambda$. However, it is sometimes
convenient to omit the subscript $\lambda$ in various symbols
for simplicity of notation.

\begin{figure}[htb]
\begin{center}
\epsfig{file=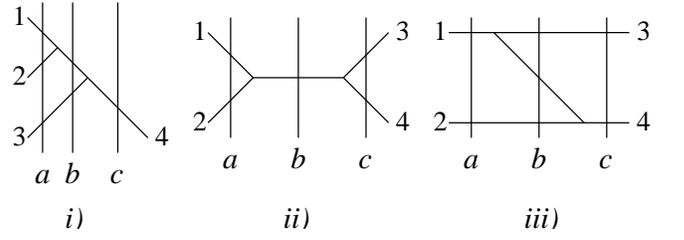}
\end{center}
\caption{Graphical illustration of interaction terms with factors
{\it i})
${\cal F}^{31}_{2\lambda}(1234)$;  {\it ii}) ${\cal
F}^{22s}_{2\lambda}(1234)$; {\it iii}) ${\cal F}^{22ex}_{2\lambda}(1234)$.}
\label{fig4}
\end{figure}

One introduces
\begin{eqnarray}
\tau_1 & = & \alpha_{21} +  \alpha_{12}, \\
\tau_2 & = & \beta_{11} + \beta_{31} + \beta_{13} + \beta_{22}
\end{eqnarray}
$\alpha_{21}$ denotes the part of $\tau_1$ that contains two creation
operators and one annihilation operator. Analogous convention is used
in all subscripts.
Equation (\ref{tau2}) gives,
\begin{eqnarray}
\beta_{\lambda 11} & = & 2 {\cal F}_{2\lambda}
[\alpha_{12}\alpha_{21}]_{11}  + \beta_{\infty 11} \label{b11} \; , \\
\beta_{\lambda 31} & = & 2 {\cal F}_{2\lambda}
[\alpha_{21}\alpha_{21}]_{31} \; , \\
\beta_{\lambda 13} & = & 2 {\cal F}_{2\lambda}
[\alpha_{12}\alpha_{12}]_{13} \; , \\
\beta_{\lambda 22}  & = & {\cal F}_{2\lambda}
[\alpha_{21}\alpha_{12}+4\alpha_{12}\alpha_{21}]_{22} \; ,
\end{eqnarray}
where
\begin{equation}
{\cal F}_{2\lambda} = \int_\lambda^\infty ds
(\{f'_s\}f_s-f_s\{f'_s\}) \; .
\end{equation}
For $f_\lambda =\exp(-ab^2/\lambda^4)$, see \cite{DEG},
\begin{equation}
{\cal F}_{2\lambda}(a,b,c)  =
\frac{p^+_{ba}ba+p^+_{bc}bc}{ba^2+bc^2}(f_{ab}f_{bc}-1) \; ,
\end{equation}
where $ab={\cal  M}^2_{a}-{\cal M}^2_{b}$   and $a$ and
$b$ are the corresponding configurations of particle momenta.
In Section III and Appendix C, factors ${\cal
F}^{22s}_{2\lambda}(1234)$, ${\cal F}^{22ex}_{2\lambda}(1234)$ and
${\cal F}^{31}_{2\lambda}(1234)$ correspond to the configurations
shown in Fig. \ref{fig4}.
\noindent Functions $V^{22}_\lambda$ and $V^{31}_\lambda$ in
Eq. (\ref{pl2}) for $P^-_{\lambda 2}$, are given in terms of
${\cal F}_{2\lambda}$ by relations:
\begin{eqnarray}
V^{22}_\lambda & = & \frac{1}{4p^+_{12}} {\cal F}^{22s}_{2\lambda} +
\frac{\theta(p^+_1-p^+_3)}{p^+_1-p^+_3} {\cal F}^{22ex}_{2\lambda} \;
, \label{V22} \\
V^{31}_\lambda & = & \frac{1}{2p^+_{12}} {\cal F}^{31}_{2\lambda} \;
. \label{V31}
\end{eqnarray}
Functions $\tilde{V}^{j22}_\lambda$ and $\tilde{V}^{j31}_\lambda$ in
Eq. (\ref{ml2}) for $M^{-j}_{2\lambda}$, are given by:
\begin{eqnarray}
\tilde{V}^{j22}_\lambda & = & \left. \frac{\theta(p^+_1-p^+_3)}{p^+_1-p^+_3}
\frac{\partial }{\partial p^j} (f_{2p}f_{3p}) \right|_{p=p_1-p_3} \; ,
\label{tV22} \\
\tilde{V}^{j31}_\lambda & = & \left. \frac{1}{2p^+_{12}} f_{12}
\frac{\partial}{\partial p^j} f_{3p}
\right|_{p=p_1+p_2} \; . \label{tV31}
\end{eqnarray}
The subscripts $ij$ of the form factors $f_{ij}$  indicate
that one of the invariant masses in the definition (\ref{deff})
is given by $(p_i + p_j)^2$, while the other by $m^2$.

Second order terms involve mass renormalization in
$\beta_{\lambda 11}$ that equals
\begin{equation}
\beta_{\lambda 11} = \int [p] \frac{\delta m^2_\lambda}{p^+}
a_p^\dagger a_p \; ,
\end{equation}
where, according to (\ref{b11}),
\begin{equation}
\delta m^2_\lambda = \delta m^2_\infty + \frac{1}{32\pi^2} \int_0^1 dx
\int_{{\cal  M}^2_0}^\infty  d {\cal  M}^2 \frac{f^2_\lambda -1}{{\cal
M}^2 - m^2}  r^2_\Delta \; . \label{ml}
\end{equation}
\noindent The counterterm contribution, $\delta m^2_\infty$, should
remove  $\Delta$ dependence from the Hamiltonian in the limit $\Delta
\rightarrow \infty$. Hence,
\begin{equation}
\delta m^2_\infty = \frac{1}{32 \pi^2} \ln \frac{\Delta}{m} + c \;.
\end{equation}
\noindent The arbitrary constant $c$ is a finite part of the
counterterm  and may be chosen in such a way that for some value of
$\lambda$, say  $\lambda_0$, mass squared in the effective Hamiltonian
equals  $m^2 + g^2 \delta m^2_0$. The value of $\delta m^2_0$ is
determined by comparing theoretical predictions with experimental
data. Then,
\begin{equation}
\delta m^2_\infty = \delta m^2_0 - \frac{1}{32\pi^2} \int_0^1 dx
\int_{{\cal  M}^2_0}^\infty d {\cal  M}^2 \frac{f^2_{\lambda_0}
-1}{{\cal  M}^2 - m^2}  r^2_\Delta \; . \label{mn}
\end{equation}
\noindent
Combining Eqs. (\ref{ml}) and (\ref{mn}), and taking the limit  $\Delta
\rightarrow \infty$, one obtains
\begin{equation}
\delta m^2_\lambda = \delta m^2_0 + \frac{1}{32\pi^2} \int_0^1 dx
\int_{{\cal  M}^2_0}^\infty d {\cal  M}^2  \frac{f^2_\lambda
-f^2_{\lambda_0}}{{\cal  M}^2 - m^2}  \; .
\end{equation}

\section{Connection between bare and effective particles}
\label{app3}

\begin{figure}[htb]
\begin{center}
\epsfig{file=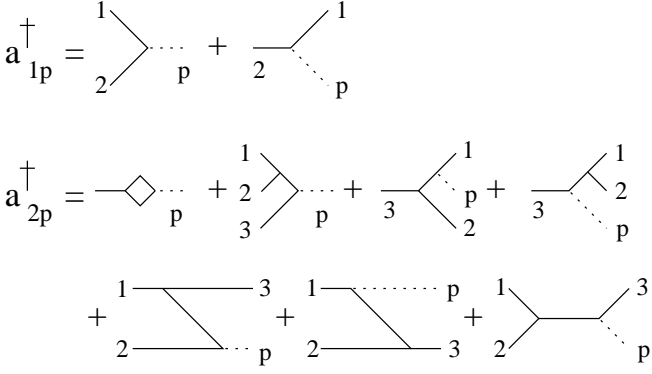}
\end{center}
\caption{Numbered lines stand for creation and annihilation operators
of effective particles, and dots indicate how momentum $p$ enters
Eqs. (\ref{a1s}) and (\ref{a2s}).}
\label{fig5}
\end{figure}

It follows from Eqs. (\ref{q}) and (\ref{dG}) that
\begin{equation}
\frac{d}{d\lambda}{\cal U}_\lambda = {\cal U}_\lambda
\frac{d}{d\lambda} \left\{(1-f){\cal G}_{I\lambda} \right\}_{{\cal
G}_0} \; . \label{dU}
\end{equation}
The initial condition is provided at $\lambda = \infty$:
${\cal U}_\infty = 1$.
Assuming an infinitesimally small $g$, ${\cal U}_\lambda$ may
be found using series expansion in powers of $g$, since
${\cal U}_\lambda (g=0)=1$, and ${\cal G}_{I\lambda}$ is made
of terms of order $g^n$ with $n \ge 1$. So,
\begin{equation}
{\cal U}_\lambda = 1+g u_{\lambda 1} + g^2 u_{\lambda 2} + \cdots \; .
\end{equation}
Equation (\ref{dU}) provides
\begin{equation}
u_{\lambda 1}= \{ (1-f_\lambda){\cal G}_{1} \} \; ,
\end{equation}
and
\begin{equation}
u_{\lambda 2} =  \frac{1}{2} u_{\lambda 1}^2 + v_{\lambda 2} \; ,
\end{equation}
where
\begin{equation}
v_{\lambda 2}= \{ (1-f_\lambda){\cal G}_{2}\}
+ \frac{1}{2} \int_\infty^\lambda ds [u_{s 1}, u'_{s 1}]  \; .
\end{equation}
Using this solution for ${\cal U}_\lambda$ and Eq. (\ref{q}),
one can write $q_\infty$ in terms of $q_\lambda$ as follows.
\begin{eqnarray}
q_{0} & = & q_\lambda \; . \\
q_{1} & = & [q_\lambda, u_{\lambda 1}] \; , \\
q_{2} & = & [q_\lambda, v_{\lambda 2}]
+ \frac{1}{2}[[q_\lambda,u_{\lambda 1}],u_{\lambda 1}] \; .
\end{eqnarray}
In full detail, see Fig. \ref{fig5},
\begin{align}
a^\dagger_{1 p} = \int [12] & \left[ 
-\frac{1}{2}\tilde{\delta} (p-p_{12}) r_{12} \frac{1-f_{12}}{p^- -
p^-_{12}}    a_{\lambda 1}^\dagger a_{\lambda 2}^\dagger \right. \nonumber \\
& \left. +\tilde{\delta} (p_2-p_{1p}) r_{1p}  
\frac{1-f_{1p}}{p^-_2 - p^-_{1p}}
a_{\lambda 2}^\dagger a_{\lambda 1} \right] \label{a1s}
\end{align}
and
\begin{widetext}
\begin{eqnarray}
a^\dagger_{2 p} & = & - \frac{1}{64\pi^2} \int_0^1 dx
\int_{{\cal  M}^2_0}^\infty d {\cal  M}^2
\frac{(1-f_{\lambda})^2}{({\cal  M}^2 - m^2)^2}  r^2_\Delta
\, a_{\lambda p}^\dagger \nonumber \\
& & - \int [123] \left\{ \frac{1}{2}
r_{12} r_{(12)3}  \frac{1}{p^+_{12}} a_+^{31}(123p)
\, \tilde{\delta} (p-p_{123}) \,
a_{\lambda 1}^\dagger a_{\lambda 2}^\dagger a_{\lambda3}^\dagger
\right. \nonumber \\
& & - \left[ r_{1p} r_{(1p)2}
\frac{1}{p^+_{1p}} \, a_-^{31}(p123)
+ \frac{1}{2} r_{12} r_{(12)p} \frac{1}{p^+_{12}}
\, a_+^{31}(12p3) \right] \tilde{\delta} (p_3-p_{12p}) \,
a_{\lambda 3}^\dagger a_{\lambda 1} a_{\lambda 2} \nonumber \\
& & + \left[ r_{3(1-3)} r_{2(p-2)}
\frac{\theta(p^+_1-p^+_3)}{p^+_1-p^+_3} \, a_+^{22ex}(123p)
+ r_{p(1-p)} r_{2(3-2)} \frac{\theta(p^+_1-p^+)}{p^+_1-p^+}
\, a_-^{22ex}(12p3) \right. \nonumber \\
& & \left. + \; r_{12} r_{3p} \frac{1}{2p^+_{12}} \, a_+^{22s}(123p) \right]
\tilde{\delta}(p_{3p}-p_{12}) \,
a_{\lambda1}^\dagger a_{\lambda 2}^\dagger a_{\lambda 3} 
\left.  \right\} \; , \label{a2s}
\end{eqnarray}
\end{widetext}
where
\begin{equation}
a_\pm= {\cal B} \pm {\cal C} \; .
\end{equation}
${\cal B}$ comes from $v_{\lambda 2}$ and ${\cal C}$ from
$1/2 [[q_\lambda,u_{\lambda 1}],u_{\lambda 1}]$.
\begin{align}
& {\cal B} = p^+_{ac} \frac{1-f_{ac}}{ca} {\cal F}_2 (a,b,c) \nonumber \\
& - \frac{1}{2} \frac{p^+_{ab}p^+_{bc}}{ba \, bc} \left[ f_{ab} - f_{bc}
-\frac{ba^2-bc^2}{ba^2+bc^2}(f_{ab} f_{bc} -1)\right] \; .
\end{align}
\begin{equation}
{\cal C} = \frac{1}{2} \frac{p^+_{ab}p^+_{bc}}{ba \, bc} 
(1- f_{ab})(1 - f_{bc}) \; .
\end{equation}
Indices in parenthesis in $r_{i(jk)}$, or $r_{i(j-k)}$, mean that 
$\kappa^\perp$ is calculated as a relative momentum of particle 
$i$ and the second particle with momentum $p_j^{+,\perp}+p_k^{+,\perp}$, 
or $p_j^{+,\perp}-p_k^{+,\perp}$, respectively. The arguments of 
$a_\pm(1234)$ refer to particle momenta in the configurations $a$, 
$b$ and $c$, which are labeled using the same convention as used 
for ${\cal F}_{2\lambda}$ in Fig. \ref{fig4}.

Note that if the operator trajectories $a_\lambda$ and Poincar\'e
generators at one finite value of $\lambda$ were known, there would be
no need to trace the changes of the generators all the way back to 
a local theory. Unfortunately, they are not known and the construction
developed here has to draw on the formal structure that a local theory implies
through perturbation theory. An exact solution of Eq. (\ref{dG}), if
it existed, could extend the approach beyond perturbation theory. Virtually
nothing is known about existence of such solutions, but if they existed,
one should expect them to possess a rich mathematical structure.

\section{Dynamical rotations}
\label{app4}

This Appendix describes derivation and features of spatial rotation
operators in LF dynamics on example of the rotation around one
of the transverse axes. The operator $U(\alpha)$ for rotations around
axis number 1 by angle $\alpha$, is calculated similarly to the $S$-matrix
in old-fashioned perturbation theory, except that the angle $\alpha$ is analogous
to a finite interval of time. One writes $U(\alpha)$ in the form
\begin{equation}
U(\alpha)=W(\alpha) e^{-i\alpha J^1_{0\lambda}} \; ,
\end{equation}
so that
\begin{equation}
W(\alpha)= e^{-i\alpha J^1_\lambda} e^{i\alpha J^1_{0\lambda}} \; .
\end{equation}
$W(\alpha)$ satisfies the differential equation
\begin{equation}
\frac{d}{d\alpha}W(\alpha)= -i W(\alpha) J^1_{I\lambda} (\alpha) \; ,
\label{W}
\end{equation}
where
\begin{equation}
J^1_{I\lambda} (\alpha)=e^{-i\alpha J^1_{0\lambda}}  J^1_{I\lambda}
e^{i\alpha J^1_{0\lambda}} \; ,
\end{equation}
and $J^1_{I\lambda} = J^1_\lambda - J^1_{0\lambda} = (M^{-2}_\lambda -
M^{-2}_{0\lambda})/2$.  When the interaction is absent,
$U(\alpha)=e^{-i\alpha J^1_{0\lambda}}$ and $W(\alpha)=1$. For
infinitesimally small coupling constants $g$, one can integrate
Eq.(\ref{W}) term by term in a power series in $g$. One obtains
\begin{equation}
W(\alpha)=1 + g W_1(\alpha) + g^2 W_2(\alpha) + \cdots \; ,
\end{equation}
where
\begin{equation}
W_1(\alpha) = -i\int_0^\alpha d\beta J^1_{\lambda 1} (\beta) \label{W1} \; ,
\end{equation}
\begin{align}
& W_2(\alpha) = W^{(11)}_2(\alpha) + W^{(2)}_2(\alpha) \nonumber\\
& = (-i)^2\int_0^\alpha d\beta \int_0^\beta d\beta' J^1_{\lambda 1}
(\beta') J^1_{\lambda 1} (\beta)  -i\int_0^\alpha d\beta
J^1_{\lambda 2} (\beta) \; .
\end{align}

\subsection{Terms independent of interactions}
Using relations
\begin{eqnarray}
e^{-i\alpha J^1_{0\lambda}} a^\dagger_{\lambda p} e^{i\alpha J^1_{0\lambda}}
& = & a^\dagger_{\lambda p_{\alpha}} \; ,\\
e^{-i\alpha J^1_{0\lambda}} a_{\lambda p} e^{i\alpha J^1_{0\lambda}}
& = & a_{\lambda p_{\alpha}} \; ,
\end{eqnarray}
one obtains
\begin{align}
& |{\Psi_\alpha}_0 (p) \rangle = U_0(\alpha)|\Psi_0(p) \rangle \nonumber\\
& = U_0(\alpha) a^\dagger_{\lambda p} U_0^{-1}(\alpha) U_0(\alpha)|0 \rangle
= a^\dagger_{\lambda p_{\alpha}} |0 \rangle = |\Psi_0 (p_{\alpha})\rangle \; ,
\end{align}
where
\begin{eqnarray}
p^+_{\alpha} & = & p^+ \cos^2 \frac{\alpha}{2}  +\frac{p^{\perp2}+m^2}{p^+}
\sin^2 \frac{\alpha}{2} + p^2 \sin \alpha \; , \nonumber \\
p^2_{\alpha} & = & p^2 \cos \alpha - \frac{1}{2} \left( p^+ -
\frac{p^{\perp2}+m^2}{p^+} \right) \sin \alpha \; , \nonumber \\
p^1_{\alpha} & = & p^1 \; . \label{p0}
\end{eqnarray}

\subsection{First order terms}

\newcommand{\si}{{\rm s}} \newcommand{\co}{{\rm c}}

According to (\ref{W1}),
\begin{equation}
W_1(\alpha)= -\frac{i}{2} \int_0^\alpha d \beta M^{-2}_{\lambda 1}
(\beta)  \; , \label{w1def}
\end{equation}
where
\begin{equation}
M^{-2}_{\lambda 1} (\beta) = \frac{i}{2} \int [p_1p_2p_3] f_\lambda
\left(  \frac{\partial}{\partial p_3^2} \tilde{\delta} \right)
a^\dagger_{1\beta} a^\dagger_{2 \beta} a_{3\beta} + h.c. \; .
\end{equation}
The subscript $\lambda$ in creation and annihilation operators
is omitted in the above formula to make it more readable.
The transformation of momenta
is given by Eq. (\ref{p0}) with $\alpha$ replaced by $\beta$.
Changing variables from $p_1$, $p_2$, $p_3$ to  $k_1=p_{1\beta}$,
$k_2=p_{2\beta}$, $k_3=p_{3\beta}$, one obtains
\begin{widetext}
\begin{equation}
M^{-2}_{\lambda 1} (\beta)
= \frac{i}{2} \int [k_1k_2k_3] f_\lambda \left[
\frac{\partial}{\partial p_3^2}
\tilde{\delta}(k_{3 \hat \beta}-k_{1 \hat \beta 2 \hat \beta }) \right]
a^\dagger_{\lambda 1} a^\dagger_{\lambda 2} a_{\lambda 3} + h.c. \; , \label{mprod}
\end{equation}
where the subscript $\hat \beta$ denotes the interaction-independent
transformation of momenta $k^+$ and $k^\perp$ that corresponds to
inverse rotation to the rotation by the angle $\beta$.
\begin{equation}
\frac{\partial}{\partial p_3^2} = \frac{1}{- k^2_3 \si + k^+_3
\tilde{\co}^2 + k^-_3 \tilde{\si}^2} \left[ \left( k^+_3
\tilde{\co}^2 - k^-_3 \tilde{\si}^2 \right) \frac{\partial}{\partial
k_3^2} + 2 \tilde{\si}  \left( k^+_3 \tilde{\co} - k^2_3 \tilde{\si}
\right) \frac{\partial}{\partial k_3^+} \right] \; , \label{p2k}
\end{equation}
where $\si \equiv \sin \beta$, $\tilde{\si} \equiv \sin \beta /2$,
$\co \equiv \cos \beta$, $\tilde{\co} \equiv \cos \beta /2$.

Note that the derivative $\partial / \partial p^2_3$, in Eq.
(\ref{mprod}), acts only on the $\delta$-function factor,
$\delta (k^2_{3 \hat \beta}-
k^2_{1 \hat \beta 2 \hat \beta})$. Therefore,
expression (\ref{p2k}) for $\partial / \partial p_3^2$ inside (\ref{mprod})
is equivalent to
\begin{equation}
\frac{\partial}{\partial p_3^2} = \frac{1}{-(k^2_3-k^2_{12}) \si +
\frac{1}{2} (k^+_3-k^+_{12}-k^-_3+k^-_{12})\co}  \frac{d}{d \beta} \; .
\end{equation}
The denominator may be simplified thanks to the factor
$\delta (k^+_{3\hat \beta}-k^+_{1 \hat \beta 2\hat \beta})$, which
implies that
$-(k^2_3-k^2_{12}) \si + (k^+_3-k^+_{12}-k^-_3+k^-_{12})\co /2
=-(k^+_3+k^-_3-k^+_{12}-k^-_{12})/2$. This way the dependence on
$\beta$ is shown to be fully contained in the arguments of the
$\delta$-functions, and
\begin{align}
\frac{d}{d \beta}
& \left[ \delta(k^2_{3\hat \beta}-k^2_{1 \hat \beta 2\hat \beta}) \right]
\delta(k^+_{3\hat \beta}-k^+_{1 \hat \beta 2 \hat \beta}) \nonumber\\
& = \frac{d}{d \beta} \left[ \delta(k^2_{3\hat \beta}-k^2_{1 \hat \beta 2\hat
\beta}) \; \delta(k^+_{3\hat \beta}-k^+_{1 \hat \beta 2\hat \beta}) \right]
- \delta(k^2_{3\hat \beta}-k^2_{1 \hat \beta 2\hat \beta})
\frac{d}{d \beta} \left[
\delta(k^+_{3\hat \beta}-k^+_{1 \hat \beta 2\hat \beta}) \right] \; .
\end{align}
The second term on the right-hand side in the above equation gives no
contribution because $d [ \delta (k^+_{3\hat \beta}-
k^+_{1 \hat \beta 2\hat \beta}) ]/ d \beta$ is proportional
to the argument of $\delta(k^2_{3\hat \beta}-k^2_{1 \hat \beta
2\hat \beta})$. So, the
differentiation over $\beta$ may be extended to the entire factor
$\tilde{\delta}(k_{3\hat \beta}-k_{1 \hat \beta 2 \hat \beta})$.

It follows that,
\begin{equation}
M^{-2}_{\lambda 1} (\beta) = - i \frac{d}{d \beta} \int [k_1k_2k_3]
\frac{f_\lambda}{k^-_{3 \hat \beta}-k^-_{1 \hat \beta 2\hat \beta}}
\tilde{\delta}(k_{3\hat \beta}-k_{1 \hat \beta 2\hat \beta})
a^\dagger_{\lambda 1} a^\dagger_{\lambda 2} a_{\lambda 3} + h.c. \; ,
\end{equation}
and one can integrate (\ref{w1def}) to obtain
\begin{equation}
W_1(\alpha)= -\frac{1}{2} \int [123] f_\lambda
\left[\frac{\tilde{\delta}(p_{3 \hat \alpha}
-p_{1 \hat \alpha 2 \hat \alpha})}{p^-_{3 \hat \alpha}
- p^-_{1 \hat \alpha 2 \hat \alpha}}
- \frac{\tilde{\delta}(p_3-p_{12})}{p^-_3-p^-_{12}} \right]
a^\dagger_{\lambda 1} a^\dagger_{\lambda 2} a_{\lambda 3} - h.c. \; .
\end{equation}
\end{widetext}
Then, the first order result for the rotated state reads
\begin{eqnarray}
|{\Psi_\alpha}_1(p) \rangle = U_1(\alpha) |\Psi_0(p)\rangle +  U_0(\alpha)
|\Psi_1(p)\rangle \nonumber \\
= \frac{1}{2} \int[12] \frac{f_\lambda}{p^-_{\alpha}-p^-_{12}}
\tilde{\delta}(p_{\alpha}-p_{12}) |12\rangle \; ,
\end{eqnarray}
as it should be.

\subsection{Second order terms}
\label{sot}

The calculation of $|{\Psi_\alpha}_2(p) \rangle$ produces two
parts.  The first paragraph below lists the three-effective-particle
contribution to the rotated state. The second paragraph shows
one-effective-particle contribution, which includes renormalization.

\subsubsection{Three-particle contributions}

The part of perturbative rotation operator $U(\alpha)$, which contributes to the
three-particle component of the rotated state $|\Psi_\alpha(p) \rangle$ of
order $g^2$, contains products of three creators and one annihilator for effective
particles. For these terms, integration in $W_2 (\alpha)$ is carried
out in steps similar to the case of $W_1 (\alpha)$ described earlier.
\begin{equation}
W_2(\alpha) = W_2^{(2)}(\alpha) + W_2^{(11)}(\alpha) \; ,
\end{equation}
\begin{widetext}
where
\begin{eqnarray}
W_2^{(2)}(\alpha) & = & -\frac{i}{2} \int_0^\alpha d \beta M^{-2}_{\lambda
2} (\beta) \nonumber \\
& = & - \int[1234] f_\lambda \left\{
\frac{\tilde{\delta}(p_{4\hat \alpha} - p_{1 \hat \alpha 2 \hat \alpha
3\hat \alpha})
}{p^-_{4\hat \alpha}-p^-_{1 \hat \alpha
2 \hat \alpha 3\hat \alpha}} V^{31}_\lambda (\hat \alpha)
- \frac{ \tilde{\delta}(p_4 - p_{123})}
{p^-_4-p^-_{123}} V^{31}_\lambda
\right\}  a_{\lambda 1}^\dagger a_{\lambda 2}^\dagger a_{\lambda 3}^\dagger
a_{\lambda 4} \nonumber \\
& & + \frac{1}{2} \int_0^\alpha d \beta \int[1234]
\frac{\tilde{\delta}(p_{4\hat \beta} - p_{1 \hat \beta 2 \hat \beta
3 \hat \beta})}
{p^-_{4\hat \beta}-p^-_{1 \hat \beta 2 \hat \beta 3 \hat \beta }}
\tilde{V}^{j31}_\lambda(\hat \beta) a_{\lambda 1}^\dagger
a_{\lambda 2}^\dagger a_{\lambda 3}^\dagger a_{\lambda 4} \; ,\label{w22}
\end{eqnarray}
\begin{eqnarray}
W_2^{(11)}(\alpha) & = & -\frac{i}{2} \int_0^\alpha d \beta W_1 (\beta)
M^{-2}_{\lambda 1} (\beta) \nonumber \\
& = & \frac{1}{2} \int[1234]
\frac{f_{12}}{m^2 - {\cal M}^2_{12}} \left\{
\frac{\tilde{\delta}(p_{4\hat \alpha} - p_{1 \hat \alpha 2 \hat \alpha
3 \hat \alpha})}
{p^-_{4\hat \alpha}-p^-_{1 \hat \alpha 2 \hat \alpha 3\hat \alpha}}
f_{(1\hat \alpha 2\hat \alpha)3 \hat \alpha}
- \frac{\tilde{\delta}(p_4 - p_{123})}{p^-_4-p^-_{123}} f_{(12)3}
\right. \nonumber \\
& & - \left. \frac{\tilde{\delta}(p_{4\hat \alpha}
- p_{(12) \hat \alpha 3\hat \alpha})}{p^-_4-p^-_{(12)\hat \alpha3\hat \alpha}}
f_{(12)3}
+ \frac{\tilde{\delta}(p_4 - p_{123})}{p^-_4-p^-_{(12)3}} f_{(12)3}
\right\}  a_{\lambda 1}^\dagger a_{\lambda 2}^\dagger a_{\lambda 3}^\dagger
a_{\lambda 4} \nonumber \\
& & - \; \frac{1}{2} \int_0^\alpha d \beta \int[1234]
\frac{\tilde{\delta}(p_{4\hat \beta} - p_{1 \hat \beta 2 \hat \beta
3 \hat \beta})}{p^-_{4\hat \beta}-p^-_{1 \hat \beta 2 \hat \beta 3 \hat \beta}}
\frac{f_{12}}{m^2 - {\cal M}^2_{12}}
\left[ \frac{d}{d \beta} f_{(1\hat \beta 2\hat \beta)3 \hat \beta} \right]
a_{\lambda 1}^\dagger
a_{\lambda 2}^\dagger a_{\lambda 3}^\dagger a_{\lambda 4} \; .\nonumber \\
\label{w211}
\end{eqnarray}
The last term in $W_2^{(2)}(\alpha)$ in Eq. (\ref{w22}) cancels the last
term in $W_2^{(11)}(\alpha)$ in Eq. (\ref{w211}).
Collecting all contributions to $|{\Psi_\alpha}_2(p) \rangle$,  one
obtains
\begin{equation}
|{\Psi_\alpha}_2 (p) \rangle_{tree} =
\Big\{ [U_2^{(2)}(\alpha) + U_2^{(11)}(\alpha)]|\Psi_0(p) \rangle
+ U_1(\alpha)|\Psi_1(p) \rangle
+ U_0(\alpha) |\Psi_2 (p) \rangle \Big\}_{tree} =
|\Psi_2(p_\alpha) \rangle\; . \label{psitree}
\end{equation}

\subsubsection{Renormalization of one-particle component}

Contributions to the one effective particle component of
$|{\Psi_\alpha}_2 (p) \rangle$, are
\begin{eqnarray}
|{\Psi_\alpha}_2 (p) \rangle_{loop} & = & \Big\{ [U_2^{(2)}(\alpha) +
U_2^{(11)}(\alpha)]|\Psi_0(p) \rangle  + U_1(\alpha)|\Psi_1(p) \rangle
\Big\}_{loop} \; , \\
\Big\{
U_2^{(2)}(\alpha) |\Psi_0(p) \rangle \Big\}_{loop} & = &
\frac{\delta m^2_\lambda}{p^+_{\alpha}} \left(
\frac{\sin \alpha}{2} \frac{\partial}{\partial p^2}
- \sin^2 \frac{\alpha}{2} \frac{\partial}{\partial p^+}  \right)
a^\dagger_{\lambda p_{\alpha}} | 0 \rangle \; , \\
\Big\{ U_2^{(11)}(\alpha) |\Psi_0(p) \rangle \Big\}_{loop}
& = & - \frac{\delta \tilde{m}^2_\lambda}{p^+_{\alpha}} \left(
\frac{\sin \alpha}{2} \frac{\partial}{\partial p^2}
- \sin^2 \frac{\alpha}{2} \frac{\partial}{\partial p^+} \right)
a^\dagger_{\lambda p_{\alpha}} |0\rangle \nonumber \\
& & + \int [123] f^2_\lambda \,
\frac{\tilde{\delta}(p_3-p_{12})}{2(p^-_3-p^-_{12})}
\left[\frac{\tilde{\delta} (p-p_{1 \hat \alpha 2\hat \alpha})}
{p^--p^-_{1 \hat \alpha 2\hat \alpha}}
-\frac{\tilde{\delta}(p_\alpha - p_{12})}
{p^-_\alpha-p^-_{12}}\right] a^\dagger_{\lambda 3} |0 \rangle \; , 
\nonumber \\
\label{u211psi0} \\
\Big\{ U_1 (\alpha) |\Psi_1 (p) \rangle  \Big\}_{loop} 
& = & \int [123] f^2_\lambda \,
\frac{\tilde{\delta}(p-p_{1 \hat \alpha 2 \hat \alpha})}
{2(p^--p^-_{1 \hat \alpha 2 \hat \alpha})}
\left[\frac{\tilde{\delta}(p_{3\hat \alpha}-p_{1 \hat \alpha 2 \hat \alpha})}
{p^-_{3\hat \alpha}-p^-_{1 \hat \alpha 2\hat \alpha}}
- \frac{\tilde{\delta} (p_3-p_{12})}{p^-_3-p^-_{12}}\right]
a^\dagger_{\lambda 3} |0\rangle \; ,\nonumber \\
\label{u1psi1}
\end{eqnarray}
\end{widetext}
where
\begin{equation}
\delta \tilde{m}^2_\lambda = \frac{1}{32\pi^2} \int_0^1 dx \int_{{\cal
M}^2_0}^\infty d {\cal  M}^2 \frac{f^2_\lambda}{{\cal M}^2 - m^2} \; . 
\label{dtm}
\end{equation}
Combining these expressions, ($\{ U_1 (\alpha) |\Psi_1 (p) \rangle \}_{loop}$,
Eq. (\ref{u1psi1}), cancels the last term in Eq. (\ref{u211psi0})), one obtains
\begin{align}
& |{\Psi_\alpha}_2 (p) \rangle_{loop} \nonumber\\
& = \frac{\delta m^2_\lambda - \delta \tilde{m}^2_\lambda}{p^+_{\alpha}}
\left( \frac{\sin \alpha}{2} \frac{\partial}{\partial p^2}
- \sin^2 \frac{\alpha}{2} \frac{\partial}{\partial p^+} \right)
a^\dagger_{\lambda p_{\alpha}} | 0 \rangle \; . \label{psiloop}
\end{align}
The complete second order term in Eq. (\ref{pa2}) is given by
the sum of the tree term with three effective particles, Eq. (\ref{psitree}),
and the loop term with mass renormalization in the term with
one effective particle, Eq. (\ref{psiloop}).

\begin{equation}
|{\Psi_\alpha}_2 (p) \rangle = |{\Psi_\alpha}_2 (p) \rangle_{tree}
+ |{\Psi_\alpha}_2 (p) \rangle_{loop} \; ,
\end{equation}
which proves (\ref{tpop}) for terms of order $g^2$.


\begin{thebibliography}{99}

\bibitem{Dirac} P. A. M. Dirac, Rev. Mod. Phys. {\bf 21}, 392 (1949).

\bibitem{Wilson12} K. Wilson, Phys. Rev. B{\bf 140}, 445 (1965);
K. G. Wilson, Phys. Rev. D{\bf 2}, 1438 (1970).

\bibitem{DEG} St. D. G{\l}azek, Phys. Rev. D{\bf 63}, 116006 (2001);
and references therein.

\bibitem{GW} St. D. G{\l}azek and K. G. Wilson, Phys. Rev. D{\bf 48},
5863 (1993); {\it ibid.} D{\bf 49}, 4214 (1994).

\bibitem{KS} See e.g. J. Kogut, L. Susskind, Phys. Rev. C{\bf 8}, 75 (1973);
H. Leutwyler and J. Stern, Ann. of Phys. {\bf 112}, 94 (1978);
B.L.G. Bakker, L.A. Kondratyuk and M.V. Terent'ev, Nucl. Phys. B{\bf 158} (1979) 497;
F. Coester and W. N. Polyzou, Phys. Rev. D{\bf 26}, 1348 (1982).

\bibitem{Yan} See e.g. S.-J. Chang, R. G. Root and T.-M. Yan
Phys. Rev. D{\bf 7}, 1133 (1973).

\bibitem{EW} E. Wigner, Ann. of Math. {\bf 40}, 149 (1939).

\bibitem{Hari} A. Harindranath, A. Mukherjee, R. Ratabole,
Phys. Rev. D{\bf 63} 045006 (2001).

\bibitem{Wilsonetal} K. G. Wilson et al., Phys.  Rev.  D{\bf 49}, 6720
(1994).

\bibitem{LB} G. P. Lepage, S. J. Brodsky, Phys.  Rev.  D{\bf 22}, 2157
(1980).

\bibitem{BPA} M. Brisudov\'a and R. J.  Perry, Phys.  Rev.  D{\bf 54},
1831 (1996); B. H. Allen, R. J. Perry, Phys.  Rev. D{\bf 62}, 025005 (2000).

\end{thebibliography}
\end{document}